\documentclass[twocolumn]{aastex6}
\pdfoutput=1

\received{May 5, 2016}
\revised{July 29, August 28, October 11, 2016}
\accepted{October 11, 2016}
\shorttitle{L1544 Magnetic Field: I}
\shortauthors{Clemens et al.}

\begin{document}

\title{The Magnetic Field of L1544: \\
I. Near-Infrared Polarimetry and the Non-Uniform Envelope}

\author{Dan P. Clemens,}
\affil{Institute for Astrophysical Research, Boston University,
    725 Commonwealth Ave, Boston, MA 02215}
\author{K. Tassis,}
\affil{Department of Physics and ITCP, University of Crete, 71003, Heraklion, Greece}
\affil{IESL, Foundation for Research and Technology-Hellas, PO Box 1527, 71110 Heraklion, Crete, Greece} 
\author{and Paul F. Goldsmith}
\affil{Jet Propulsion Laboratory, M/S 169-504, 4800 Oak Grove Drive,
Pasadena, CA 91109}
\email{clemens@bu.edu, tassis@physics.uoc.gr, paul.f.goldsmith@jpl.nasa.gov}

\begin{abstract}
The magnetic field (B-field) of the starless dark cloud L1544 has been studied using 
near-infrared (NIR) background starlight polarimetry (BSP) and archival data in order 
to characterize the properties of the plane-of-sky B-field. NIR linear polarization 
measurements of over 1,700 stars were obtained in the $H$ band and
201 of these were also measured in the $K$ band. The NIR BSP properties are correlated with 
reddening, as traced using the RJCE 
($H$ - $M$) method, and with thermal dust emission from the L1544 cloud and envelope seen
in {\it Herschel} maps. The NIR polarization position angles change at the
location of the cloud and exhibit their lowest dispersion of position angles there,
offering strong evidence that NIR polarization traces the plane-of-sky B-field of L1544.
In this paper, the uniformity of the plane-of-sky B-field in the envelope region of L1544 is quantitatively assessed. This allowed evaluating the approach of assuming uniform field geometry when measuring relative mass-to-flux ratios in the cloud envelope and core based on 
averaging of the envelope radio Zeeman observations, as in \citet{Crutcher09}.
In L1544, the NIR BSP shows the envelope B-field to be 
significantly non-uniform and likely not suitable for
averaging Zeeman properties without treating intrinsic variations.
Deeper analyses of the NIR BSP and related data sets, 
including estimates of the B-field strength and testing how it varies with position and
gas density, are the subjects of later papers in this series.
\end{abstract}

\keywords{Galaxy: disk -- ISM: magnetic fields -- magnetic fields -- ISM: individual: 
(L1544) -- polarization -- techniques: polarimetry}

\section{Introduction}

Magnetic fields (B-fields) are present in the diffuse interstellar material from which 
dark, molecular clouds form. B-fields are also present 
in the cores of those clouds, some of which form new stars. 
Theoretical modeling of cloud and star formation in the presence of magnetic 
fields has a long, rich history \citet{Mestel56} which continues through recent work \citep[e.g.,][]{Mouschovias91, Galli93, Basu94, Padoan99, Mouschovias06, Hennebelle08, Li15} and comprehensive reviews \citep[e.g.,][]{Mouschovias96a,Mouschovias96b}. Observational tests of such theories have been rarer \citep[e.g.,][]{Goodman89, Goodman95, Crutcher99, Crutcher09, Zhang14, Pillai15}, but are now advancing at an accelerated pace \citep[see review by][]{Li14}.
A key quantity used to assess cloud stability and predict future outcomes
is the ratio of mass, in the form of gas and dust, to the flux of the B-field threading
that material. This `mass-to-flux' ratio 
\citep[e.g.,][]{Mestel56, Mouschovias76a, Mouschovias76b, Fiedler92, Crutcher12} is 
normally indexed by its critical ratio, the value at which the gravitational and
magnetic energy densities are equal, yielding a $M / \Phi$ ratio of unity. If $M / \Phi$ is sub-unity,
then B-fields dominate and the region is classified as sub-critical. If $M / \Phi$ exceeds 
unity, the region is super-critical, with B-fields overwhelmed by gravity, leading to contraction or 
collapse. The differential $M / \Phi$ of a cloud core relative to its envelope is a key
indicator of the role of B-fields in star formation \citep{Mouschovias76a}.

However, measuring $M / \Phi$ is challenging. Assessing the numerator involves sensing
atomic or molecular gas densities, temperatures, and columns. This is made especially
difficult when depletion onto dust grains robs the gas of the already rare species used as 
proxy tracers for the dominant H$_2$ gas. Tracing gas by using dust, as revealed 
through the reddening and extinction
of starlight (or background diffuse emission) or via thermal emission from the dust
grains, is less affected by depletion, though not fully immune. And, depending on the
collision rate of gas and dust, the dust temperature may
not closely reflect the gas temperature. Well-sampled, large-area
gas spectral line maps, sensitive multi-wavelength dust maps, and multi-species, 
multi-line analyses over cloud envelope and core
regions, where properties change rapidly with location, are necessary to address 
these difficulties.

Computing the denominator in $M / \Phi$ is
even more difficult. The B-field is a three-dimensional vector field, yet current best
methods can only probe either the line-of-sight component, $B_{LOS}$, 
employing the Zeeman effect for radio spectral
line observations, or the plane-of-sky component, $B_{POS}$, using background starlight polarimetry (BSP) or linear polarization of the thermal dust emission. An alternate spectral line
method, exploiting the Goldreich-Kylafis (1981) effect, may return more than single dimension
information, but only in special anisotropic settings \citep[e.g.,][]{Girart99}.

Radio Zeeman observations need significant gas column densities and relatively quiescent 
conditions to yield detectable signals within reasonable integration times. Hence, the number
of targets observed using the Zeeman effect for the OH molecule, which is best for the typical
conditions in molecular clouds \citep[while CN is better for massive, star-forming 
cloud cores;][]{Crutcher96}, is not large. The number of OH Zeeman
effect {\it detections} is even smaller. Yet, the Zeeman effect directly returns 
the line of sight component of the B-field strength, making it a `gold standard' tool
for assessing cloud stability.

Polarized thermal dust emission in the mm, submm, and far-infrared (FIR) 
reveals the orientations of $B_{POS}$, 
but only where the emission (and the weaker polarization)
can be detected. This favors the densest and warmest cloud core locations. 
\citet{Crutcher04} compared the $B_{POS}$ amplitude, inferred from the 
\citet{WardThompson00} James Clerk Maxwell Telescope (JCMT) SCUBA 
\citep{Holland99} 850~$\mu$m polarization map, with the \citet{Crutcher00}
OH Zeeman $B_{LOS}$ amplitude for L1544. That comparison indicated 
more than an order of magnitude discrepancy. \citet{Crutcher04}
attributed this to the different sizes of the structures traced by the two methods and
to the dependence of field strength on density under flux-freezing conditions.
A better approach, one that samples the same structures and densities as the Zeeman 
OH method, is needed.

Background starlight polarimetry, especially performed in the dust-penetrating 
near-infrared (NIR), reliably reveals
B-field orientations in the plane-of-the-sky, and can do so with higher angular resolution
and numbers of directions probed than available Zeeman observations.
But, using BSP to develop B-field strengths, and so assess $M / \Phi$ ratios and 
cloud and core stability, depends on a longer and more complex chain of arguments. Nevertheless, the ease of obtaining BSP B-field maps offers the opportunity to gain insight into
B-field properties over a remarkably wide range of diffuse and molecular cloud conditions.

\subsection{The L1544 Laboratory}

L1544 represents a nearly ideal laboratory for comparing the Zeeman and BSP
methods. It is a molecular cloud
with a starless dense cloud core \citep{Snell81, Myers83a, Heyer87} in Taurus, at a distance 
of about 140~pc \citep{Elias78, Kenyon94, Torres12}. Detection of ammonia by \citet{Myers83b} 
and N$_2$H$^+$, C$_3$H$_2$, and CCS by \citet{Benson98} began a long history of 
interstellar chemistry and kinematic studies of L1544, as it supports
an unusually chemical richness \citep[e.g.,][]{Caselli02a, Lee03, vanderTak05, Vastel06, 
Bizzocchi14}, strong depletion of 
many species onto cold dust grains \citep{Tafalla02, Keto10}, and rotation plus infall motions 
\citep{Tafalla98, Williams99, Williams06, Keto15}. 
L1544 was the first dark cloud core to be detected in the fundamental ortho-water transition
using {\it Herschel} \citep{Caselli10} and continues to serve as a key laboratory for modeling
water in dark clouds \citep{Keto14}.  Detailed modeling of the physical conditions, stability, 
ionization state, and magnetic properties of L1544 have been pursued in several studies 
\citep[e.g.,][]{Ciolek00, Tafalla02, Li02a, Li02b, Crapsi05, Keto08}. 

The strong depletion of CO isotopologues and the resultant uncertainty regarding the
abundances of these species have favored observations of optically thin dust continuum
emission for revealing the physical nature of the L1544 core and envelope regions. 
Studies include
1.3~mm mapping at IRAM by \citet{WardThompson99}, 850~$\mu$m mapping
by \citet{Shirley00} and \citet{WardThompson00}, and 450~$\mu$m mapping 
\citep{Shirley00, Doty05}.
The {\it Herschel} satellite was used to map L1544 at 70, 100, 160, 250, 350, 
and 500~$\mu$m wavelengths in programs by O. Krause and P. Andr\'{e} (obs IDs
1342193503/4 and 1342204841/2, respectively). These sensitive images reveal 
the L1544 cloud core and envelope with fine detail. Figure 1 shows the 250~$\mu$m
{\it Herschel}/SPIRE map as the background image, with black contours stepped 
linearly with 250~$\mu$m intensity. 

% F1
\begin{figure}[h]
\plotone{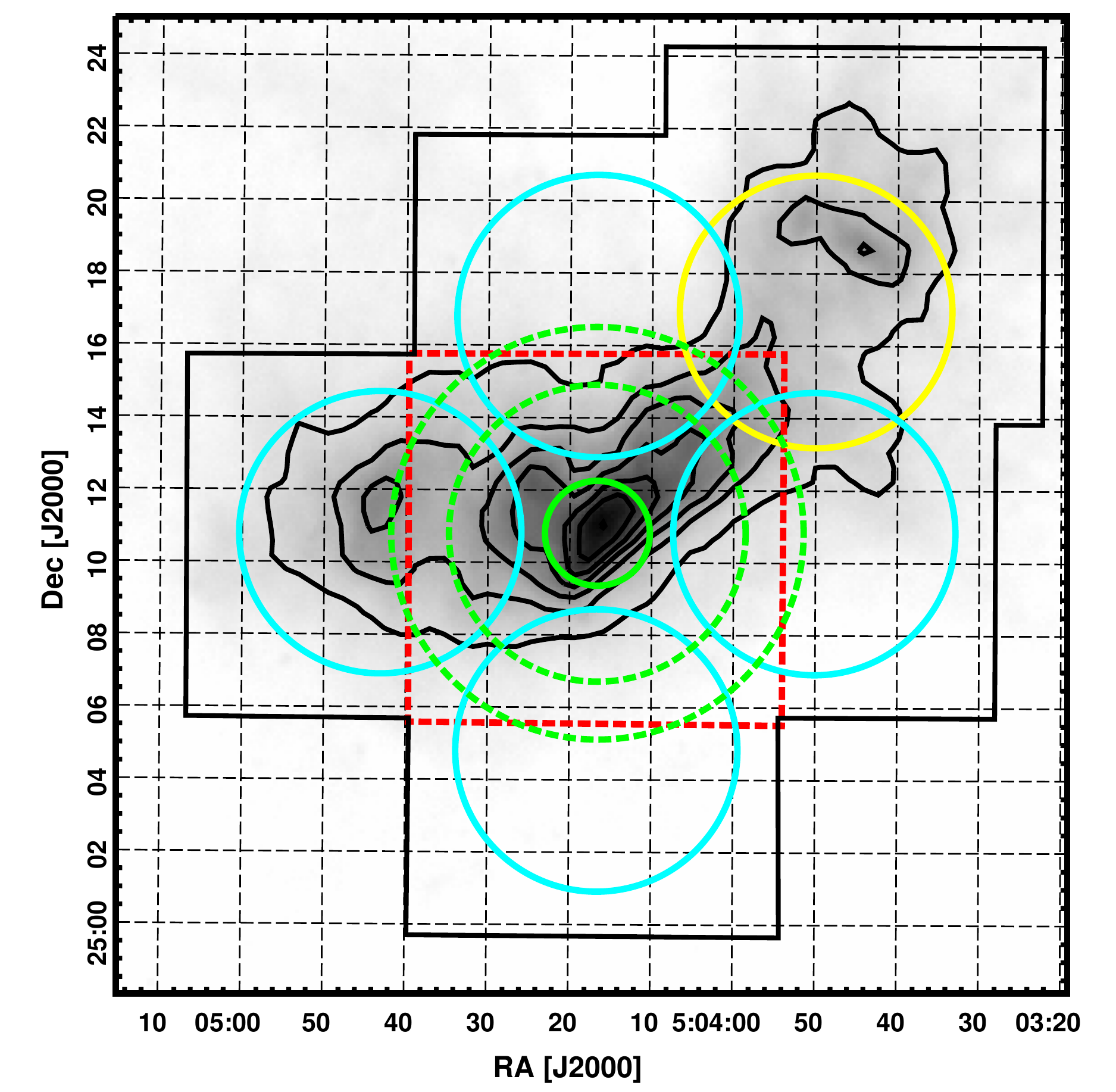}
\caption{Gray-scale and contour map of {\it Herschel}/SPIRE 250~$\mu$m dust emission
covering L1544. The dense cloud core is seen as the elongated, mostly gray to 
black structure in the figure center. Black contours start at 75~MJy~sr$^{-1}$ and increase 
in steps of 25~MJy~sr$^{-1}$.
The solid green circle and dashed green circles identify the Arecibo main beam
size (solid) and first sidelobe region (dashed) for the OH Zeeman observations of \citet{Crutcher09}.
Four cyan circles identify the placements and beam size of their GBT OH Zeeman observations.
The yellow circle identifies the placement and beam size of new Effelsberg OH Zeeman 
observations (K. Tassis 2016, private communication).
The black polygon identifies the region surveyed for NIR $H$ band 
polarization while the red dashed box identifies the region surveyed for $K$ band polarization, 
both using Mimir in this current study.
\label{fig_Herschel}}
\end{figure}

The dust found in the dense cores of dark clouds like L1544 is optically thick at optical and 
near-infrared wavelengths and optically thin in the mid-infrared \citep{Bacmann00}. 
The optically thin core of 
L1544 is easily seen as resolved, faint absorption in {\it WISE} \citep{Wright10} W3 
(12~$\mu$m) and W4 (22~$\mu$m) maps of the region, which show the same orientation
and structure exhibited by the bright FIR emission core of Figure~\ref{fig_Herschel}.

The L1544 B-field strength was first measured via detection of the OH Zeeman effect
by \citet{Crutcher00}, who found $B_{LOS} = +11 \pm 2$~$\mu$G using the Arecibo telescope
along the direction to the cloud core in a nearly 16~hour integration. In the Arecibo OH Zeeman
effect survey of 34 dark cloud cores \citep{Troland08}, L1544 exhibits the second
highest Zeeman signal-to-noise ratio (SNR), at 6.4. The fraction of survey targets with SNR 
$\ge$ 3 is only 21\% and these used about 38\% of the total integration time of the survey, 
underscoring the difficulty of determining B-field strengths from radio Zeeman observations.

To test predictions of ambipolar diffusion (AD) models \citep[e.g.,][]{Ciolek94} for cloud core 
development leading to star formation, \citet{Crutcher09} combined
Arecibo OH Zeeman observations (beamsize $\sim 2.9$~arcmin) of four cloud cores with 
GBT OH Zeeman observations (beamsize $\sim 7.8$~arcmin) of their associated 
cloud core envelopes. The latter measurements were performed by sampling four 
positions located outside the Arecibo beam observation of each cloud core, with 
offsets along each equatorial cardinal direction of $\pm 6$~arcmin. The goal was to 
test whether the cloud cores and envelopes exhibited relative $M / \Phi$ ratios in excess of unity, 
as predicted by AD models.  Of the four clouds observed, only a single 
envelope position toward L1544 exhibited a 
GBT Zeeman detection with SNR~$\ge$~3. \citet{Crutcher09} averaged the four envelope
GBT Zeeman values for each cloud to increase the SNR and thereby permit estimating 
or limiting envelope $M / \Phi$
values. However, this approach was criticized by \citet{Mouschovias10} on the basis of
likely non-uniformity of the envelope B-fields over the angular extent of the regions sampled
by the GBT beams, based on the appearance of the {\it Herschel} and {\it Spitzer} 
dust thermal emission distributions on
the sky for L1544 and the other three dark clouds.

Submm dust emission polarization at 850~$\mu$m was sought in the L1544 core
by \citet{WardThompson00}, who used JCMT with SCUBA and its
polarimetry module \citep[][hereafter `SCUPOL']{Greaves03}. They obtained detections 
toward eight
Nyquist-sampled directions, all within the \citet{Crutcher09} Arecibo core OH Zeeman beam. 
One star (identified as 2MASS PSC J05041591+251157) just outside the core OH Zeeman beam
was measured for $K$-band polarization by \citet{Jones15}.

Based on the condensed, quiescent, starless nature of its dense core, the
presence of a resolved envelope, and the two, independent OH Zeeman detections of
B-fields (toward the core and toward one envelope position), L1544 appears to be 
an ideal laboratory for conducting
deep, NIR BSP observations and for the analyses of the resulting B-field orientations.
Heretofore, no independent probe of the envelope B-fields has
been reported for any of the four dark clouds studied using the OH Zeeman effect
by \citet{Crutcher09}. The NIR BSP project described here was performed to directly
addresses the nature of the envelope B-field in L1544.

\subsection{Goals and Methodology}

The first project goal was to obtain NIR BSP detections in regions that fully covered the 
portions of
L1544 observed by the Arecibo and GBT OH Zeeman observations, to enable the desired
polarization and B-field comparisons. NIR wavelengths (Section~\ref{sec:observations}) 
offer good sensitivity for polarization
detections toward regions of moderate dust extinction (A$_V \sim 1-2$~mag) up 
to the much higher values
(20 -- 30 mag) characterizing the outer parts of dense cloud cores, through which optical
starlight cannot penetrate. This observational goal was met through use of 
the Mimir NIR imaging polarimeter \citep{Clemens07} to deeply probe the L1544 core and
to less deeply probe five fields offset from the core, four along the same cardinal directions
as those of the GBT OH observations \citep{Crutcher09}, and one to cover the region examined
in recent radio Zeeman OH observations conducted at Effelsberg (K. Tassis 2016, private
communication). 
BSP was performed over all these
fields in the NIR $H$-band~(1.6~$\mu$m) and also toward the center field in the
NIR $K$-band~(2.2~$\mu$m). These observations and the data reduction details are described
in Section 2, below.

The second project goal was to demonstrate that the NIR polarizations returned by the
Mimir observations revealed the B-field associated with L1544 and its core. Such
association of BSP and molecular cloud B-fields has been challenged by \citet{Arce98}
as being due only to a `skin depth' effect, though \citet{Whittet08} found
little evidence of such an effect. In Section~\ref{sec:analysis}, descriptions are 
presented of two tests that were performed 
to confirm the association of BSP and the B-field of L1544. The first
test showed that NIR BSP is correlated with stellar 
reddening for the same stars and that this reddening is correlated with the dust thermal
emission from L1544. The second test showed that the BSP-traced B-field properties
of polarization position angle, PA, and dispersion of polarization position angle, $\Delta_{PA}$, 
exhibit strong spatial  correspondences with the L1544 dust maps. 

Section~\ref{sec:discussion} presents an assessment of the mean BSP-traced $B_{POS}$ 
properties measured in synthetic beam averages representing the Zeeman radio beams 
(the same angular sizes and directions) used by \citet{Crutcher09}. This included 
analyzing possible 
non-uniformity of the envelope $B_{POS}$. This has impact on a key assumption underlying 
the \citet{Crutcher09} envelope-averaging of Zeeman data, on their resulting conclusions regarding $M / \Phi$, and thus on their test of ambipolar diffusion. 

In later papers in this series, the NIR BSP data are 
used to develop a map of BSP-traced B-field strength. This map is
compared to the Arecibo and GBT OH Zeeman-traced B-field strengths of \citet{Crutcher09},
as well as new OH Zeeman observations conducted at the 100~m Effelsberg 
telescope to assess the degree of correspondence between the two techniques. 
A map of $M / \Phi$ across L1544 is also used to reassess the \citet{Crutcher09} 
conclusions regarding AD in L1544. 

\section{Observations and Data Processing}
\label{sec:observations}

Near-infrared imaging polarimetric observations of L1544 were conducted
on the UT nights of 2013 January 20, 2015 October 27 and 31, 2015 November 1 and 2,
and 2016 January 27 and 31 using the Mimir instrument
\citep{Clemens07} on the 1.83~m Perkins Telescope, located outside Flagstaff, AZ.
Mimir polarimetry used cold, rotated, compound half-wave plates (HWP) for modulating
the polarization signal for each of the $H$ and $K$
wavebands and a fixed wire-grid analyzer preceding the $1024 \times 1024$~pixel InSb 
ALADDIN III detector array. All polarization and reimaging optics in Mimir operated
at 60-70~K and the detector array was at 33.5~K. The plate scale was 0.58~arcsec
per pixel, resulting in a $10 \times 10$~arcmin field of view (FOV). All observations
were conducted through fewer than 1.4 airmasses and the seeing was better than
2~arcsec for all observations.

The observations, in each of the two bands, were performed by obtaining an image
through a fixed angular orientation of the HWP then rotating the HWP to other angles and
obtaining additional images. A total of 16 HWP angle-images, chosen to permit forming
four independent sets of Stokes $U$ and $Q$ parameters for each star, were observed at each
telescope pointing. The telescope performed a set of six 
sky pointings (as a rotated hex pattern), with 16 HWP angle-images obtained at 
each pointing, resulting in 96 images per observation. 

Six pointing centers (`fields') were selected toward a Center direction 
($\alpha = 05^h$04$^m$16$^s$.6, 
$\delta = +$25$\degr$10$^\prime$48$^{\prime\prime}$ [J2000]), 
the four equatorial NSEW directions offset by 6 arcmin from the Center, and
one field to the NW offset mostly diagonally by about 10.9 arcmin.  
In the $H$ band, the observation
sets included one dithered observation, at 2.5~s per exposure, toward each of the
five fields. The four NSEW fields were also observed with four longer (15~s per 
exposure) polarimetric observations in the $H$ band. The NW field had eight 15~s per exposure
$H$ band polarimetric observations. The Center field had one 2.5~s per exposure
$K$ band observation, two 15~s per 
exposure $K$ band observations, an additional 2.5~s $H$, plus four 10~s 
and eight 15~s per exposure $H$ band observations. The total integration times were thus 
about 1.7 hours in $H$ band for each NSEW field, 3.6 hours in the NW field in $H$, 
0.9 hours in $K$ toward the Center field, 
and 4.5 hours in $H$ toward the Center field. The shorter exposures served the purpose of
extending the observational dynamic range to stars whose brightness would saturate in the 
longer exposures. The longer Center and NW integration times partially offset the higher 
extinctions present in these fields.

Calibration consisted of application of detector linearity and dark current corrections, 
in-dome polarization flat-fields in each band, as well as super-sky flat-fielding using
the observed images, correction for secondary instrumental polarization across
the FOV (determined from observations of mostly unpolarized globular cluster
stars located off the Galactic plane), and HWP offset angle correction (determined
from observations of polarization standard stars). Details of the observation
methodology, data correction steps, and calibration
are described in \citet{Clemens12a,Clemens12b}.

The long (10s or 15s) observations for each band were combined to yield deep photometric
images and polarimetric point source catalogs. 
The short (2.5s) observations were also processed to polarimetric point
source catalogs. The short and long polarimetric catalogs were combined by 
matching star positions and computing variance-weighted mean $U$ and $Q$ values
and deriving from them polarization percentages, $P$, and PA values and uncertainties. 
(All $P$ values and PA uncertainties, $\sigma_{PA}$,
were corrected for the effects of positive bias in $P$, following \citet{Wardle1974}.)

The combined $H$ band polarimetric point source catalog had 1,712 entries, while the 
$K$ band catalog had 201 entries. The latter is smaller because of the smaller solid
angle observed in $K$ band (29\% of the $H$ band solid angle), the much shorter net
integration time in $K$ band (20\% of the $H$ band integration time), 
the lower mean polarizations in $K$
compared to $H$ ($\sim$ 60\%, see below) for normal ISM dust \citep{Serkowski75},
and the higher net extinction in the Center field compared to the NSEW fields
(see Figure~\ref{fig_Herschel}). The positions of all of the $K$ band stars 
matched to $H$ band stars.

In addition, the AllWISE \citep{Cutri13} catalog entries for the field shown in 
Figure~\ref{fig_Herschel}
were fetched using DS9 \citep{Joye03} to query the Centre de Données
astronomiques de Strasbourg (CDS), resulting in 3,616 stars with at least
a {\it WISE} W1 (3.6~$\mu$m) band or W2 (4.5~$\mu$m) band detection.
These were positionally matched to the $H$ band catalog, yielding 1,262 matches. 
The properties of the 450 $H$ band stars without {\it WISE} star matches and the
{\it WISE} stars without $H$ band star matches were similar in being fainter than the
subset of stars with $H$-to-{\it WISE} matches. As the fainter stars 
in the $H$ band polarimetric catalog were not expected to yield 
polarization detections providing significant B-field
information \citep{Clemens12a,Clemens12c}, this 26\% non-match rate 
among the faint stars should not bias
any findings based on the brighter stars.

\begin{splitdeluxetable}{ccCccccccBccccccccl}		% for two column
\tabletypesize{\tiny}
\tablecaption{Stellar Polarimetry and Photometry in the L1544 Field\label{tab_NIR_pols}}
\tablewidth{0cm}
\colnumbers
\tablehead{
&&\multicolumn{10}{c}{Mimir Values / Uncertainties}&\multicolumn{3}{c}{2MASS Values / Unc.}&\multicolumn{2}{c}{WISE Values / Unc.}&\\
\cline{3-12} \cline{13-15} \cline{16-17}
\colhead{No.} & \colhead{RA/decl} & \colhead{$H$} & \colhead{$P^\prime_H$} & \colhead{PA$_H$} & \colhead{$Q_H$} & \colhead{$U_H$} & \colhead{$K$} & \colhead{$P^\prime_K$} & 
\colhead{PA$_K$}& \colhead{$Q_K$} & \colhead{$U_K$} &
\colhead{$J$} & \colhead{$H$} & \colhead{$K$} & \colhead{W1} & \colhead{W2} & 
\colhead{Notes}\\
&\colhead{[$\degr$]} & \colhead{[mag]} & \colhead{[\%]} &
 \colhead{[$\degr$]} &  \colhead{[\%]} & \colhead{[\%]} & \colhead{[mag]} & \colhead{[\%]} & \colhead{[$\degr$]} &  \colhead{[\%]} & \colhead{[\%]} &
\colhead{[mag]} & \colhead{[mag]} &  \colhead{[mag]} &  \colhead{[mag]} &
 \colhead{[mag]} &
}
\startdata
0001&75.84865&15.615&  8.053& 42.6&  0.797&  9.684&20.000&  0.000&  0.0&  0.000&  0.000&16.232&15.511&15.193&15.363&15.692&\\ [-3pt]
    &25.26836& 0.037&  5.437& 19.3&  5.504&  5.437&99.999&100.000&180.0&100.000&100.000& 0.099& 0.109& 0.134& 0.047& 0.146&\\
0002&75.84951&16.313&  8.781&  8.2& 11.058&  3.262&20.000&  0.000&  0.0&  0.000&  0.000&20.000&20.000&20.000&15.796&15.857&\\ [-3pt]
    &25.36376& 0.058&  7.471& 24.4&  7.468&  7.496&99.999&100.000&180.0&100.000&100.000&99.999&99.999&99.999& 0.060& 0.162&\\ [-4pt]
$\sim$\\ [-4pt]
0455&75.98059&12.889&  1.447& 62.4& -0.840&  1.205&12.640&  2.358& 27.0&  1.718&  2.365&13.683&13.000&12.716&12.554&12.480&\\ [-3pt]
    &25.18452& 0.010&  0.249&  4.9&  0.250&  0.248& 0.017&  1.728& 21.0&  2.055&  1.527& 0.034& 0.032& 0.025& 0.024& 0.026&\\ [-4pt]
$\sim$\\ [-4pt]
0605&76.00536&11.766&  0.000&  0.0& -0.155& -0.057&20.000&  0.000&  0.0&  0.000&  0.000&12.444&11.815&11.658&11.448&11.388&\\ [-3pt]
    &25.03555& 0.001&  0.462&180.0&  0.460&  0.479&99.999&100.000&180.0&100.000&100.000& 0.045& 0.037& 0.027& 0.053& 0.055&\\
0606&76.00553&10.350&  0.142& 47.3& -0.015&  0.188&20.000&  0.000&  0.0&  0.000&  0.000&10.893&10.422&10.243&10.158&10.136&a\\ [-3pt]
    &25.03394& 0.001&  0.124& 25.1&  0.123&  0.124&99.999&100.000&180.0&100.000&100.000& 0.031& 0.036& 0.025& 0.028& 0.028&\\
\hline
\enddata
\tablecomments{This is a shortened version of the full table that is available in electronic form, with 
the rows shown here containing entries with, and without, $K$-band polarimetry and $WISE$ photometry.} 
\tablenotetext{a}{Foreground star}
\end{splitdeluxetable}

The combined data set of 1,712 stars is listed in Table~\ref{tab_NIR_pols}. 
In the Table, the first column lists the star number and the second
column presents the RA  and decl on successive lines. Columns
3 -- 7 list the Mimir-based $H$ band photometric magnitude, (debiased) 
linear polarization percentage, $P^\prime_H$, (equatorial) polarization PA$_H$ 
(in deg E from N),
and Stokes $Q_H$ and $U_H$ percentages.  Columns 8 -- 12 list the
Mimir $K$ band values of magnitude, $P^\prime_K$, PA$_K$, $Q_K$, and $U_K$ for the 201 
matching stars. Values of 20.000 in the $K$ band mag column
signify the absence of a matching star. For those stars, the corresponding
$P^\prime$ entry was set to zero, $\sigma_P$ was set to 99.99\%, and PA 
was set to zero, with $\sigma_{PA}$ set to 180.00\degr. 
Uncertainties are found in the line immediately following a line of values.

For stars matched to 2MASS \citep{Skrutskie06, Cutri03} point sources, columns
13 -- 15 provide the $J$, $H$, and $K$ band magnitudes, with their uncertainties
in the same columns on the following line.
Where no magnitude is available, a value of 20.000 and uncertainty of 9.999 are
inserted. The 1,262 {\it WISE} W1 and W2 band magnitudes are provided in 
columns 16 and 17, also following the convention that a missing magnitude is
replaced with a value of 20.000 and uncertainty 9.999. The final column lists 
letters linked to notes following the Table.

Stellar polarizations obtained using imaging surveys, as performed here, result
in wide ranges of polarimetric uncertainties, due to stellar faintness and sky
brightness, as described in \citet{Clemens12c}. Users of 
Table~\ref{tab_NIR_pols} are cautioned to consider the biases 
in such data.

\section{Analysis}
\label{sec:analysis}

\subsection{Do NIR Polarizations Reveal the L1544 B-field?} 

Establishing the nature of the B-field in L1544 from NIR BSP
rests on showing that the lines of sight to the background stars sample dust in the
L1544 cloud and core and that the polarizations originate in that dust. In the past,
BSP for other Taurus clouds has been argued to
be incapable of revealing B-fields within the clouds, but instead reveals only those B-fields
residing on the surfaces of the clouds \citep{Arce98}. In the following, the 
Table~\ref{tab_NIR_pols} data set is shown to probe the L1544 envelope and
core dust  and the B-fields there. This begins by developing
a map of stellar reddening and comparing to the {\it Herschel} thermal dust
emission of the cloud (Figure~\ref{fig_Herschel}). In the second phase, 
unique cloud physical characteristics are shown to correlate with unique polarization
properties changes - conditions unlikely to obtain unless the B-field of the cloud also
participates in those changes and is traced by the BSP up through
extinctions as high as A$_V \sim 40$~mag.

\subsubsection{Stellar Reddening Map}

The Rayleigh-Jeans Color Excess (RJCE) method of \citet{Majewski11} reveals the 
stellar reddening
caused by interstellar dust from $H$ and $M$ (4.5~$\mu$m) band
magnitude differences, as compared to
intrinsic $(H - M)$ stellar colors. RJCE is superior to similar techniques that do not use the $M$ band,
because of the narrower range of intrinsic stellar colors and because $M$ band
easily penetrates most dark clouds. The data set in Table~\ref{tab_NIR_pols} were used
to develop $(H - M)$ colors from the Mimir, 2MASS, and {\it WISE} magnitudes. 
These colors were 
interpolated to create a map of reddening, as described in what follows.

There are 772 stars in Table~\ref{tab_NIR_pols} simultaneously having 2MASS $H$ and
{\it WISE} W2 magnitude uncertainties smaller than 0.5 mag, resulting in that many $(H - M)$ 
colors with uncertainties below 0.7 mag. There are another 386 stars in the Table without
such 2MASS $H$ band data, but which have good W2 band data. Their Mimir $H$ band 
magnitudes, which are normally not color-corrected but are matched to 2MASS $H$ zero 
points in each field
\citep{Clemens12a}, can be used if color-corrected. This was implemented by finding
the zero-point $(H_{Mimir} - H_{2MASS})$ and the color term $(H_{2MASS} - W2)$
dependences on $(H_{Mimir} - W2)$. The final combined set of $(H - M)$ colors
contained 1,158 stars, distributed mostly uniformly across the region. 

This Mimir survey region area of about 420 sq arcmin, sampled by 1,158 stars, results in just
under three measured stellar reddenings per sq arcmin. Interpolation onto a $10 \times 10$~arcsec
grid included weighting each star's color by its inverse
color variance and gaussian-tapered offset from each grid center. The effective angular 
resolution of the resulting $(H - M)$ color map was set
by the stellar search radius from each grid center, the gaussian offset taper, and a 
final gaussian smoothing of the gridded interpolation. The effective number of stars 
used in the calculation of the interpolated reddening at each grid center was also found. 
At a resolution of 120 arcsec FWHM, no fewer than two stars were used to estimate
color at each grid point (this was in the most opaque part of the cloud core) and a 
mean of 10 to 11 stars characterized the grid center color estimates across the map. 

\begin{figure}
\plotone{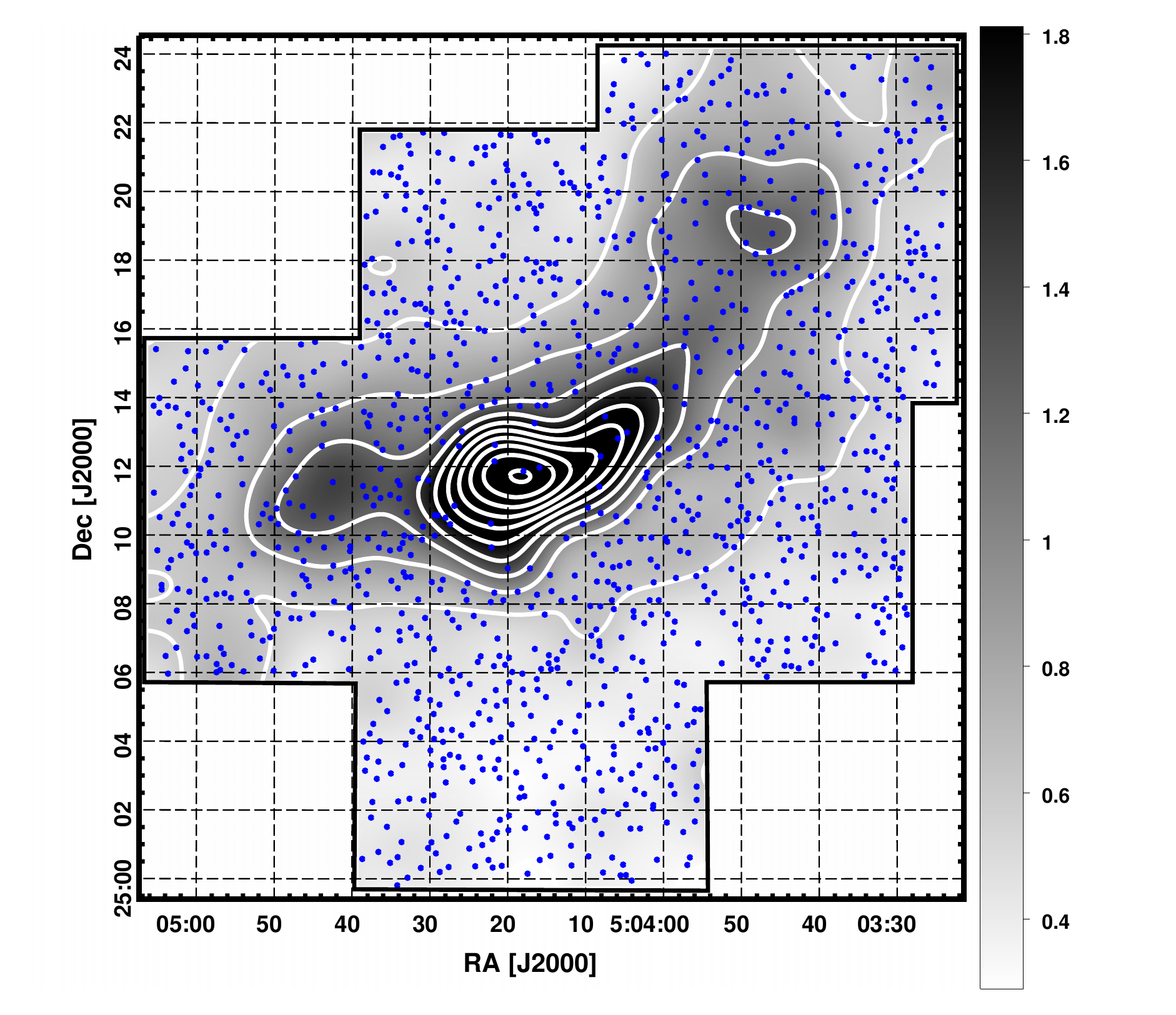}
\caption{Gray-scale NIR stellar reddening map, computed from $(H - M)$ colors
across the Mimir survey region. Grayscale wedge at right indicates mapping of darkness
to $(H - M)$ color, in magnitudes. White contours begin at 0.6 mag and are stepped by
0.3 mag to a maximum of 3.3 mag (A$_V > 25$~mag). Blue filled circles identify the locations of the 
stars used to form the interpolated map. The highest reddening is offset from the
FIR dust emission center (see Figure~\ref{fig_Herschel}) because of the dearth of 
background stars visible through that center. The outer three contours closely 
follow the {\it Herschel}\ 250~$\mu$m dust emission structure,
confirming that the emitting dust is also extincting and (as will be shown below) polarizing.
\label{fig_HM_map}}
\end{figure}

Figure~\ref{fig_HM_map} shows the resulting $(H - M)$ map. The gray-scale image fills
the Mimir survey region (outlined in black). White contours run from 0.6 through 3.3 mag
of $(H - M)$ color. The blue, filled, circles show the locations of the 1,158
stars. The reddening structure revealed in
this map closely follows the location and structure of the {\it Herschel}\  250~$\mu$m 
dust emission shown in Figure~\ref{fig_Herschel}, including the general location of the 
dense core, the lower density extension running from the core to the west-northwest 
\citep[i.e., L1544-W;][]{Tafalla98},
and the partially detached lower density feature located east of the core 
\citep[i.e., L1544-E;][]{Tafalla98}.
Thus, the BSP stellar polarization sample of Table~\ref{tab_NIR_pols} is
well-suited to probing the dust and gas structure of the L1544 dark cloud. 

% F2

Interestingly, the region of Figure~\ref{fig_HM_map}
showing $(H - M)$ colors bluer than 0.6 mag (i.e., outside the outermost contour)
still shows significant color of about
0.4 mag. For an average intrinsic $(H - M)$ color of 0.08 mag \citep{Majewski11},
the color excess is 0.32 mag, equivalent to A$_K = 0.29$ mag or 
A$_V = 2.2$~to~$2.6$~mag (depending on the extinction curve chosen, as
embodied in different $R_V$ values),
distributed throughout the region surrounding L1544. 
The greatest reddening contour in Figure~\ref{fig_HM_map} is offset from the peak
emission in Figure~\ref{fig_Herschel} for two reasons. First, the blue dot pattern shows
that there are no Table~\ref{tab_NIR_pols} stars with $(H - M)$ colors located within 
120 arcsec of the direction of the strongest FIR thermal dust emission, leaving the interpolated 
reddenings there based on the surrounding
stars. Second, the star closest to the central contour has the greatest $(H - M)$ color
of the sample, at 5.5 mag (A$_V \ge 40$~mag), affecting or biasing the contour locations to some
degree. Nevertheless, outside the very core, the stellar reddenings closely follow the dust thermal 
emission structure, giving high confidence that these stars can probe the B-field in their 
directions.

\subsubsection{Foreground Star Census}
\label{sec:foreground}

One potential concern involves the possible biasing effects of including polarization 
values for stars located in the foreground with those for stars behind the cloud.
Foreground stars in this field were sought using proper motion screening and color plus 
polarization screening. Thirteen stars exhibit proper motions in UCAC3 \citep[The Third 
U.S. Naval Observatory CCD Astrograph Catalog;][]{Zacharias10} with SNRs in 
either RA or decl exceeding 2.5. Of these, all but two have $(H - M)$ colors
that are redder than the extended region value of 0.4 mag, and were thereby judged 
to be background to the cloud. One of the survivors has $H = 16.2$~mag 
(i.e., is quite faint) and 
$\sigma_{PA}$ of 60$\degr$\, (i.e., has a poorly constrained polarization PA)
and so is unlikely to affect polarization findings no matter its 
location classification. The lone remaining
star (number 606 in Table~\ref{tab_NIR_pols}) with detectable proper motion 
($0.42 \pm 0.04$~arcsec per year) has $m_H = 10.4$~mag,
$(H - M) = 0.22$~mag, and $P = 0.14 \pm 0.12$~\%. Such low reddening and 
low polarization from a relatively bright star with measurable proper motion make it 
almost certainly a foreground star.

Other similar stars were sought in the $(H - M)$ subsample of Table~\ref{tab_NIR_pols}, through
color and polarization selections (stars bluer than some color limit and less polarized
than some polarization limit). However, all such cuts returned samples of potential foreground
stars that failed to be uniformly distributed across the Mimir survey region. In particular,
all selections resulted in trial samples that avoided the high extinction zone of L1544.
If there were significant numbers of foreground stars, some fraction of them should 
be projected against the mostly opaque core. Instead, no stars bluer than 
$(H - M)$ of 0.6 mag appear within any 0.8 mag contour in the $(H - M)$ image.

Hence, only one foreground star was detected with certainty, no significant population of
foreground stars brighter than $H \sim 14$ can be present, and foreground stars fainter
than 14th mag are likely to be only a few and will offer nearly zero contribution to the
polarization maps and interpretations thereof. As a result, corrections for foreground
extinction and polarization were deemed unnecessary.

\subsubsection{$K$-Band Polarizations}

The restricted solid angle observed for $K$-band polarization and the shorter integration
times yielded only 201 stars with Mimir-measured polarizations in this band. Yet they 
provide important checks of the nature of the B-field and dust grains along these lines of 
sight. If polarization properties changed significantly as a function of wavelength or
dust column density, reflected here in $(H - M)$ reddening, then disentangling changes
in dust properties from changes in B-field properties would be more complex. The $K$
band polarization values were therefore compared to the $H$ band polarizations values
to assess their correlation.

A subsample of the 201 $K$-band stars was selected based on requiring that $\sigma_{PA}$ 
in both $H$ and $K$ be below 30$\degr$, to select good, or
better quality data. This yielded 39 stars. For these, the variance-weighted (from propagated
uncertainties) mean band ratio of polarization percentage and the mean band
difference in position angles were
found to be $P_K / P_H = 0.62 \pm 0.04$ and 
$\mathrm{PA}_H - \mathrm{PA}_K = 8.9 \pm 1.7\degr$, 
respectively. 
The polarization ratio is higher than values expected for dust grains
obeying the Serkowski Law of polarization versus wavelength \citep{Serkowski75, 
Wilking80} and having $\lambda_{MAX}$ (the wavelength of maximum polarization)
in the range 0.3 -- 0.8~$\mu$m that is typically found \citep{Wilking80}. 
Instead, values of $\lambda_{MAX}$
in the 1.0 -- 1.2~$\mu$m range would be needed. Hence, the elevated $P_K / P_H$
ratio signifies
that grain growth has taken place. This agrees with the strong depletion of gas
phase molecules known to occur in L1544 \citep{Caselli02b}.

The position angle offset is rather less revealing. The small offset from zero
could be reflective of remnant $K$ band HWP zero angle calibration, which was based on 
far fewer observations
of many fewer standard stars than for the extensive $H$ band calibration performed
to support the GPIPS project \citep{Clemens12b}. Alternatively, it could signify small
changes in B-field and dust properties along the line-of-sight \citep{Messinger97}.

Dependence on reddening, a proxy for dust column density, was examined via polynomial trial
fits of $P_K / P_H$ versus $(H - M)$ color, and the same polynomial
trial fits for $\mathrm{PA}_H - \mathrm{PA}_K$ versus that color. F-tests showed
that no significant linear or higher terms were detectable in the 39 member stellar
subsample, and thereby likely in the remainder of the sample. Interestingly,
there is no difference in the $P_K / P_H$ ratio when the stars are split into
a high reddening ($(H - M) > 0.9$ mag) and a low reddening sample. This implies that
grain growth must have also occurred in the cloud envelope or periphery, as well 
as in the dense core.

The overall conclusion is that
dust properties along the lines of sight probed by both $H$ band polarization and
$K$ band polarization are similar. There is no evidence of major dust property changes
that would prevent B-field interpretations of the measured polarization position angle
values, this despite the evidence for larger than normal grains. 
The uncertainties in the $K$ band polarization data are significantly 
larger than the $H$ band polarization data for the same stars, so much so that
combining the PAs measured in the two bands, when weighted by their inverse
variances, are indistinguishable from the $H$ band values alone. Hence, for most of the
remainder of the analyses, only the $H$ band polarization properties are analyzed
and reported (though the $K$ band values remain in Table~\ref{tab_NIR_pols} to 
support other studies). The $K$ band data are included in the final test of $B_{POS}$ 
properties across the GBT Zeeman beams, in Section~\ref{sec:non_uniform}.

\subsubsection{Stellar PA Map}

\begin{figure}
\plotone{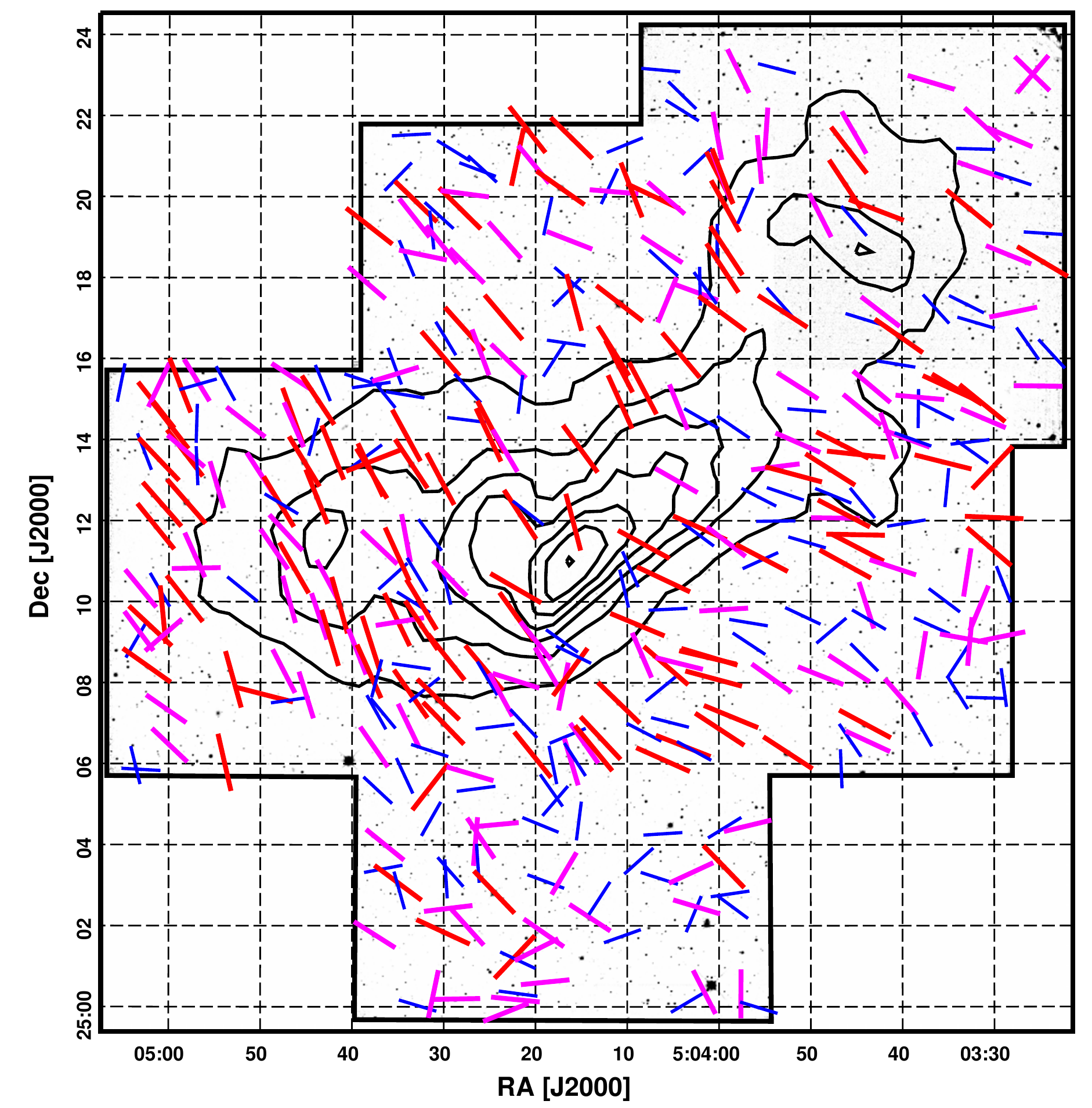}
\caption{Gray-scale $H$ band image mosaic of Mimir surveyed region. Black contours
indicate the {\it Herschel}\,  250~$\mu$m dust emission of Figure~\ref{fig_Herschel}. 
Colored line segments (`vectors') display the $H$ band PAs and $\sigma_{PA}$, for subsets 
drawn from Table~\ref{tab_NIR_pols}.
Thicker, longer, red vectors display PA orientations for the 123 stars having 
$\sigma_{PA} \le 10\degr$. Medium thick, medium length, magenta vectors are for 
the 122 stars with $\sigma_{PA}$ between $10 - 15\degr$. Thinner, shorter, blue vectors 
are for the 151 stars with
$\sigma_{PA}$ between $15 - 20\degr$. The PAs show significant star-to-star correlation
while also revealing large-scale variation in overall PA orientations across the survey region.
\label{fig_pol_vecs}}
\end{figure}

Figure~\ref{fig_pol_vecs} displays the $H$ band polarization position angles for the
396 stars in Table~\ref{tab_NIR_pols} having $\sigma_{PA} \le 20\degr$. These are 
grouped by $\sigma_{PA}$ value and coded into line 
segment (`vector') color, thickness, and length groups to better identify the highest quality values. 
The 123 stars with $\sigma_{PA} \le 10\degr$\ have longer, thicker, red colored
vectors. The middle 122 stars, with $\sigma_{PA}$ between 10 and 15$\degr$\ have
average length, average thickness, magenta colored vectors.
The 151 stars with $\sigma_{PA}$ between 15 and 20$\degr$\ have shorter, thinner, 
blue colored vectors.  Together, these 396 stars
represent the 23\% of the full sample having the lowest $\sigma_{PA}$ values. In addition
to the colored vectors, the {\it Herschel} 250~$\mu$m dust emission contours from 
Figure~\ref{fig_Herschel} are reproduced here to begin the association of NIR polarization
vector properties with dust emission properties in L1544.

% F3

The large number of vectors reveals several clear trends in
the polarization properties. First, there is a high degree of correlation of PA orientations
among neighboring stars. This basic uniformity is one of the best indicators that
a large-scale B-field is being revealed across many pc of cloud extent. Second, there is
an apparent change in mean PA across the surveyed region. In the western zone, the
PAs are more nearly horizontal ($\sim$65$\degr$\ PA) while in the eastern zone, they are 
more nearly
vertical ($\sim$25$\degr$\ PA). Hence, the mean field direction, as projected onto the plane
of the sky, is seen to change in the region of the L1544 cloud and core.
Third, the PAs seen for stars projected behind the strongest dust emission zones
are not greatly different in orientation than those seen just outside those zones. 
Fourth, the dispersion in PA values among local vectors varies with position across the
map. The eastern zone shows a high degree of star-to-star PA agreement, and thus a 
small PA dispersion, while the southern and northern zones show stronger local variations
of PAs. 

\subsubsection{NIR BSP $P$, PA, and $\Delta_{PA}$ Smoothed Images}

Spatial means of the NIR BSP values were developed to allow detailed comparison of
the L1544 B-field properties with the gas and dust distribution. As was done for the
$(H - M)$ map, weighted mean properties
were computed on 10~arcsec spaced grid centers, including all stars out to 2.3~arcmin
away from each center, and using inverse variance weighting and gaussian tapering
by each star's offset from the grid centers. This gaussian taper width ($\sigma$) was set to
1~arcmin, to favor values for stars located closest to the grid centers. 
The values computed at each grid center were
created from as few as 7 stars (at the opaque cloud core) to as many as 34 stars. The
mean number of stars per grid center was 16.6. The grid was 
smoothed with a 
second gaussian ($\sigma = 50$~arcsec), yielding images with 3~arcmin FWHM
resolution, comparable to the Arecibo Zeeman beamsize used by \citet{Crutcher09}, and
much smaller than the GBT Zeeman beamsize. 

% F4
\begin{figure}
\plotone{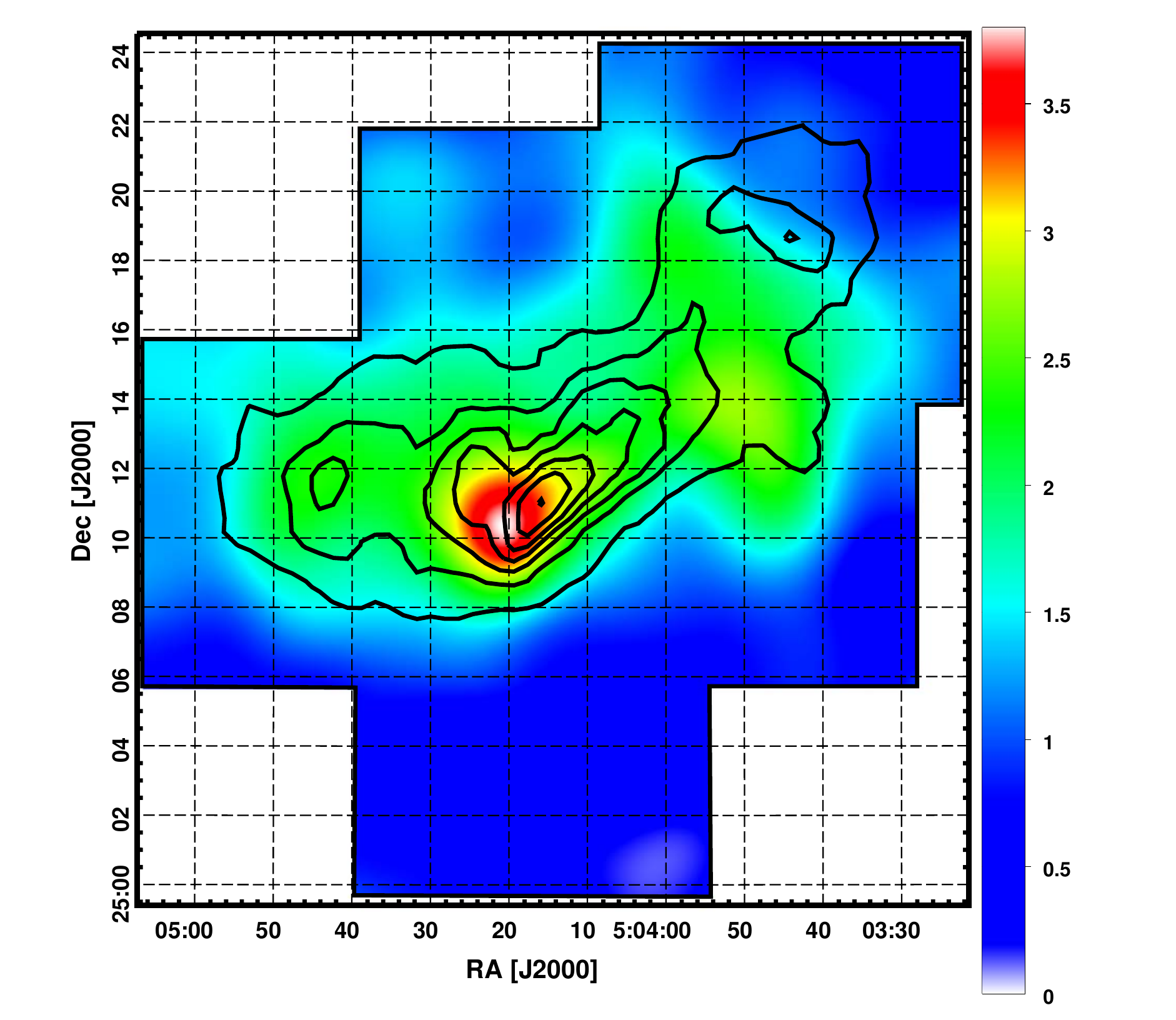}
\caption{False-color map of mean $H$ band polarization percentage, $P^\prime_H$, 
across L1544, smoothed to 3~arcmin FWHM angular 
resolution. Color look-up wedge at right indicates the mapping of $P^\prime_H$ (in percent) to displayed colors. Black contours show the
{\it Herschel}\, 250~$\mu$m emission displayed in Figures~\ref{fig_Herschel} and 
\ref{fig_pol_vecs}. 
\label{fig_mean_P}}
\end{figure}

Figure \ref{fig_mean_P} displays an image of the mean NIR $H$ band polarization 
percentage ($P^\prime_H$) distribution, and includes contours
of the 250~$\mu$m dust emission traced by {\it Herschel}. Though there are some small
differences (the NIR BSP is too extincted to probe the brightest 250~$\mu$m core emission),
overall the NIR BSP polarization percentage is significantly enhanced where L1544 dust emission 
is strongest. Thus the
NIR polarizations are tracing the B-field embedded in the same dust that
is emitting at submm wavelengths. There appear to be three spatial peaks in the $P^\prime_H$
map: a strong peak at the dust emission center; a weaker peak offset by 8~arcmin to the 
NW (South of L1544-W); and another weaker peak offset by 5.5~arcmin to the ENE (L1544-E). 

\begin{figure}
\plotone{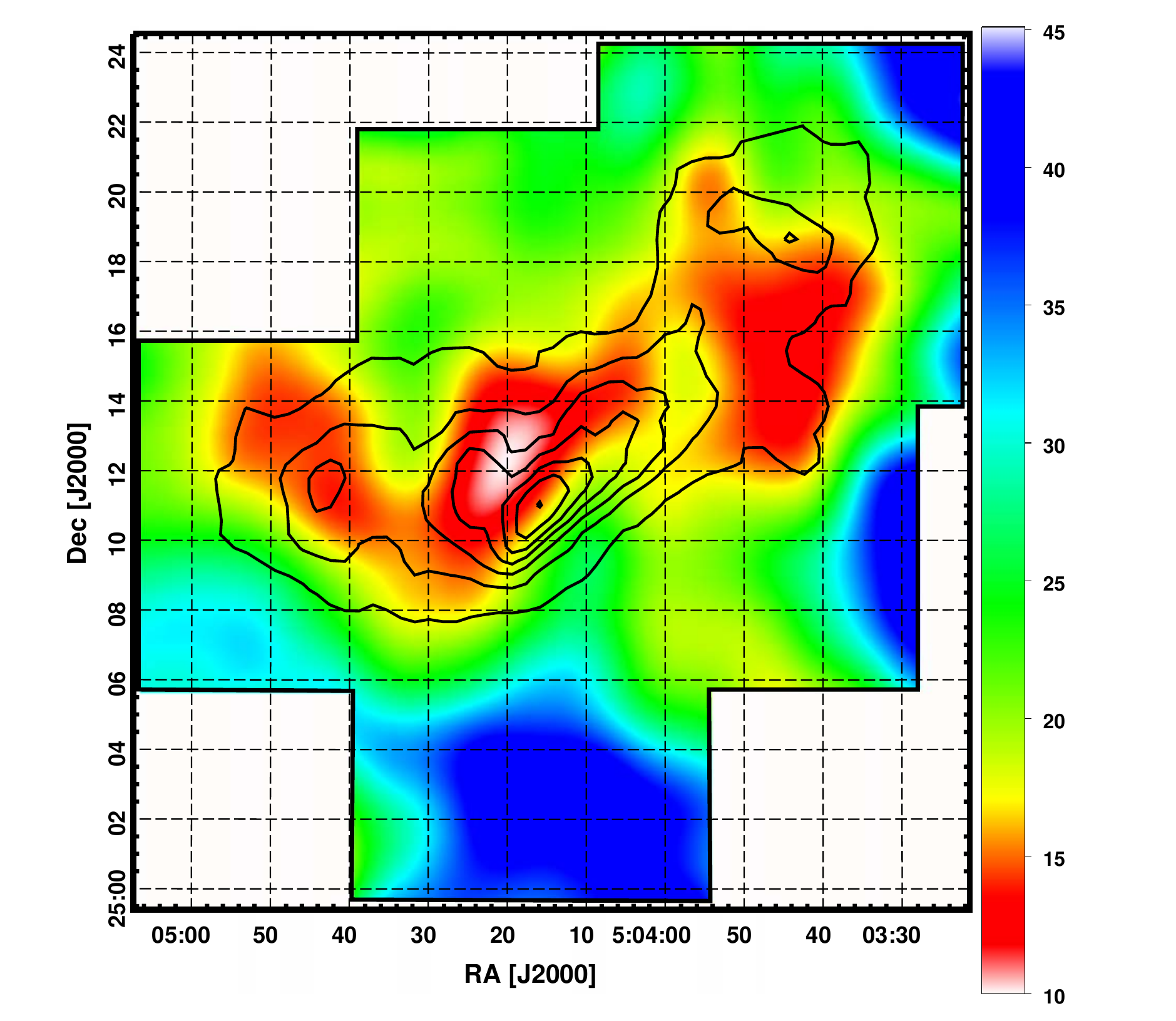}
\caption{False-color map of $H$ band polarization position angle 
dispersion, $\Delta_{PA}$ (in degrees), across L1544, at 3~arcmin 
FWHM effective angular resolution. Color look-up wedge at right indicates mapping 
of $\Delta_{PA}$ to displayed colors.
Black contours show the {\it Herschel}\, 250~$\mu$m emission. 
\label{fig_PA_disp}}
\end{figure}

The dispersion in polarization position angle, $\Delta_{PA}$, is a measure of the degree
to which background starlight reveals variations in the plane-of-sky B-field PAs. 
Figure~\ref{fig_PA_disp} displays the spatial 
distribution of $\Delta_{PA}$, compared to the {\it Herschel}\, 250~$\mu$m dust emission. 
The map was computed similarly to the $P^\prime_H$ map, measuring PA dispersions with the
same 3~arcmin FWHM resolution. This effectively removes the effects of any PA
orientation changes on angular scales larger than this, but includes in the 
computed dispersions any changes on scales smaller than 3~arcmin. Also, the biasing effects
of the individual stellar angular uncertainties in the overall PA dispersions were removed
in quadrature, using a procedure similar to that used for SCUPOL data by
\citet{Crutcher04}. In the Figure, note that the false-color scale is inverted to highlight 
where $\Delta_{PA}$ values are small, as might indicate higher B-field strengths according
to the Chandrasekhar-Fermi (\citeyear{CF53}; hereafter `C-F') method.

% F5

The minimum $\Delta_{PA}$ value is just under 10$\degr$ in the cloud core, and 
minima of 13$\degr$\, are 
seen very near the $P^\prime_H$ maxima locations to the ENE and NW. 
Thus, $P^\prime_H$ maxima and $\Delta_{PA}$ minima arise in the same
material, likely signifying where conditions are quiescent and the B-field is more uniform
and perhaps stronger.
The NIR BSP $\Delta_{PA}$ features are correlated with the {\it Herschel}\,  250~$\mu$m 
emission, though
there does appear to be somewhat of an offset along the `spine' of the cloud. The
sense of the displacement is that the minimum in $\Delta_{PA}$ is located about 2~arcmin
to the NE of the similar ridge of 250~$\mu$m dust emission. One possible cause of this
offset is revealed in the map of mean PA, below.

\begin{figure}
\plotone{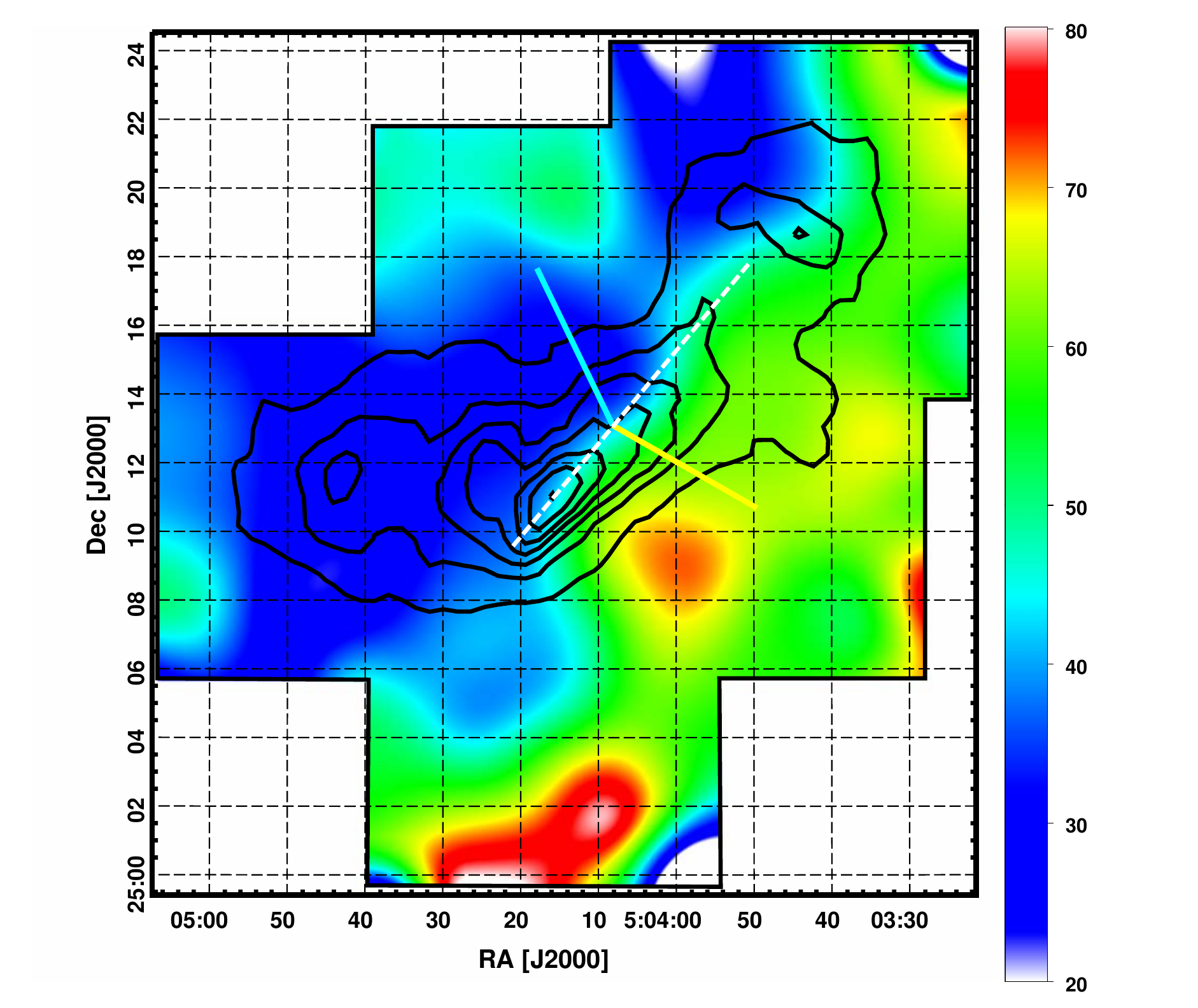}
\caption{False-color map of mean $H$ band polarization position angle, PA (in degrees),  
across L1544, at 3~arcmin FWHM effective angular resolution. 
Color look-up wedge at right indicates mapping of PA (in deg) to displayed colors.
Black contours show the {\it Herschel}\,  250~$\mu$m emission. 
The white, dashed line indicates the position angle of the cloud infrared emission `spine.'
The yellow, solid line shows the mean polarization PA South and West of the spine,
and the cyan, solid line shows the mean polarization PA North and East of the spine.
These indicate $B_{POS}$ orientations that are nearly perpendicular to the spine, but 
are also strongly changing at the position of the L1544 cloud spine.
\label{fig_mean_PA}}
\end{figure}

Figure~\ref{fig_mean_PA} displays the mean PA, computed over the same region
as for the previous two Figures. Rather than forming the mean PA directly from the BSP 
values, which introduces significant aliasing, the mean Stokes $U$ and $Q$ were 
computed from the individual stellar values
and these means were used to generate the mean PA map. The most striking finding is 
the strong gradient in mean PA coincident with the spine of L1544 infrared 
{\it Herschel} 250~$\mu$m emission, extending from the cloud
core to the NW. Along this spine, the polarization PA changes abruptly from about 
60-65$\degr$\, to 30$\degr$\, over a physical size smaller than the resolution of this
map ($\sim$0.1~pc). This swing in PA is made more interesting by comparing it to the
direction of the spine (PA$\sim -40\degr$; the white, dashed line in the Figure). 
Along the spine, the polarization PA changes from being roughly perpendicular to the cloud for directions
in the W (an 80$\degr$\, difference; the yellow, solid line in the Figure) to being similarly 
perpendicular, but with a somewhat smaller acute angle difference (70$\degr$; the cyan,
solid line in the Figure) for positions N and E of the ridge.

% F6

One possible explanation is that the cloud ridge is located at a boundary, or collision, between 
two distinct magnetic media, with different $B_{POS}$ PAs. 
Another explanation is that the B-field changes close to the dust ridge because
the B-field has significant helical pitch \citep{Fiege00} and manifests different $B_{POS}$ PAs on the two `sides' of the cloud ridge. 

\subsubsection{Gas Kinematics and B-fields}

The steep PA gradient associated with the elongated gas and dust ridge in L1544 might be
expected to be associated with a similar gradient in the radial velocity of the gas, possibly generated
through rotation of the L1544 cloud or envelope. Interestingly, while some velocity gradients
have been detected in spectroscopic maps of various gas tracers, there is no clear indication of
rotation.

A velocity gradient, mostly along the major axis of the cloud
complex, is present when comparing the radial velocity of L1544-W (10~arcmin 
offset), L1544, and L1544-E (5~arcmin offset). \citet{Heyer87} found an offset of about 0.4~km~s$^{-1}$ across 37-40~arcmin
in $^{13}$CO, yielding a gradient of 0.3~km~s$^{-1}$~pc$^{-1}$. Similarly, \citet{Tafalla98}
used C$^{18}$O to reveal a somewhat larger gradient of 1.1~km~s$^{-1}$~pc$^{-1}$
along the major axis. Their channel maps also reveal that the L1544 core exhibits another
velocity gradient, of about 3.4~km~s$^{-1}$~pc$^{-1}$, along the decl direction.
Using N$_2$H$^+$, \citet{Williams99}, \citet{Caselli02c}, and \citet{Williams06} found 
core velocity gradients of 3.8. 1.0, and 4.1~km~s$^{-1}$~pc$^{-1}$, respectively, 
with the latter two value mostly along the decl axis. 

However, neither the large-scale velocity gradient along the major cloud axis nor the smaller
scale decl velocity gradient across the L1544 core exhibit a strong correlation with the polarization
PA gradient shown in Figure~\ref{fig_mean_PA}. It may be that a weak shear due to 
the large-scale velocity gradient is bending the plane of sky B-field lines from PA 60$\degr$\, 
to PA 30$\degr$\, at the location of the cloud's major axis, but why this would affect BSP
PAs only on one side of the cloud (W) and not the other is unclear. There appears to be 
no evidence of strong cloud or core rotation and neither of the observed
weak velocity gradients appears
to explain the BSP PA change across L1544.

\subsection{SCUBA 850~$\mu$m Dust Emission Polarization in the L1544 Core}

The SCUPOL instrument combination was used on the JCMT by \citet{WardThompson00} 
to probe the L1544 core for linear polarization of the thermal dust emission at 
850~$\mu$m. These data were also used by \citet{Crutcher04} to estimate
the $B_{POS}$ field strength in the core, using the C-F
method. \citet{Matthews09} reprocessed
all SCUPOL data taken on the JCMT to produce an improved and uniformly calibrated
legacy data archive. This included refined gain calibration for all pixels and yielded improved maps
of Stokes $I$, $Q$, $U$, and their uncertainties. These reprocessed SCUPOL data for 
L1544 were obtained from the legacy archive and post-processed for this current study 
using techniques similar to those described above.

The polarization SNRs at the native $10 \times 10$~arcsec pixel sizes (the diffraction-limited
beam size was 14~arcsec) in the archive data are too low to yield adequate constraints on the
polarization position angles for more than
a couple of positions. This relates to the small number of positions (8) showing $P$ SNR$>2$
selected by \citet{WardThompson00} for plotting and used by \citet{Crutcher04} for C-F method
analysis. But, post-processing options are available which utilize more of the submm 
information and can increase the number of independent positions with detectable submm
polarization. Here, to boost the SNR (though at the expense of angular resolution), 
the Stokes $Q$ and $U$ maps from the archive were 
smoothed, using weighting by both the archive $Q$ and $U$ variance maps and a gaussian taper, 
of FWHM 35~arcsec, and resampled onto 25~arcsec pixels. From these
smoothed Stokes parameter maps, maps of $P$ and its uncertainty $\sigma_P$ were 
developed and debiased to yield $P^\prime$ values. A mask image
was generated which selected all (smoothed, resampled) map pixels having 
$P^\prime / \sigma_P \ge 1.9$, corresponding to $\sigma_{PA} \le 15\degr$, and
which had Stokes~$I$ values more than 25\% of the peak, smoothed value.
A polarization PA map was computed from the smoothed Stokes maps, 
masked, and the PA values were rotated by 90$\degr$\, to represent $B_{POS}$ orientations.

\begin{figure}
\plotone{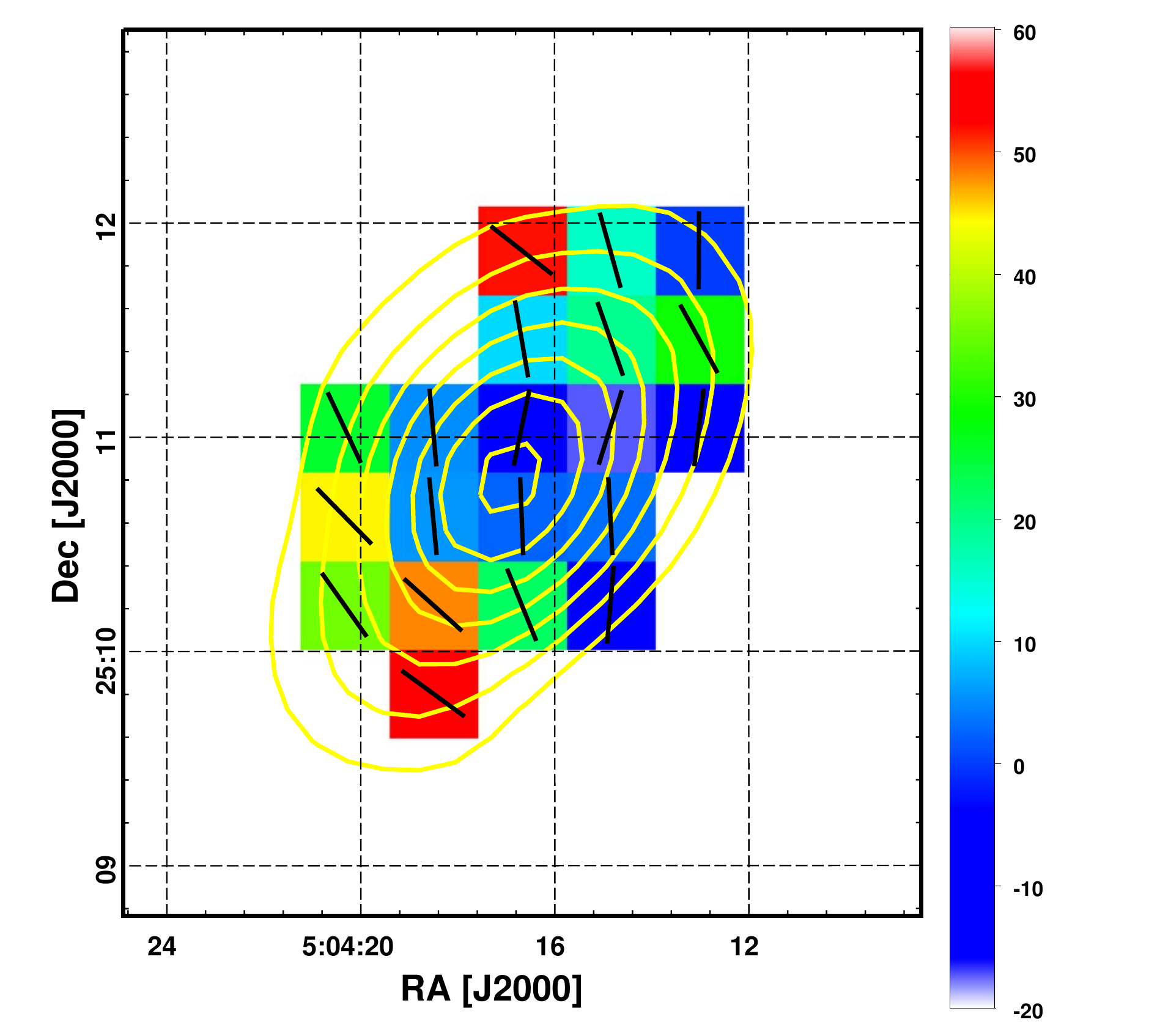}
\caption{False-color and black vector representations of the $B_{POS}$ PA (in degrees
and rotated 90$\degr$\, from the dust emission electric field polarization PA) 
measured by
SCUPOL at 850~$\mu$m, after smoothing the Stokes parameters maps to 
35~arcsec FWHM and sampling on a 25~arcsec grid. Color look up table at right 
shows how colors map to PA (in deg). Pixels shown have $\sigma_{PA} \le 15\degr$.
Yellow contours show Stokes~$I$ surface brightness at 850~$\mu$m, starting at 
43~kJy~sr$^{-1}$ and increasing in steps of half of that value. 
\label{fig_SCUBA}}
\end{figure}

Figure~\ref{fig_SCUBA} displays the twenty map pixels that passed the mask operation,
with $B_{POS}$ PA values coded into the color of each 25~arcsec pixel and the
orientations of the black line segments. 
The yellow contours show the Stokes $I$ distribution
for the 850~$\mu$m dust emission. The pixel $P^\prime$ and $B_{POS}$ PA values are listed in 
Table~\ref{tab_SCUPOL}. The variance-weighted mean PA of all 20 points is 
$9.9 \pm 1.6\degr$\, and the $\Delta_{PA}$(raw) is 20.4$\degr$. Following 
\citet{Crutcher04}, $\Delta_{PA}$(raw) was debiased by the
mean $\sigma_{PA} = 8.3\degr$\, to yield
$\Delta_{PA}$(corrected) of 18.6$\degr$. This value is larger than found by 
\citet{Crutcher04} for the eight positions reported by \citet{WardThompson00} 
for the SCUPOL data, but prior to the \citet{Matthews09} reprocessing.

% F7

% SCUPOL derived zone polarizations - basis for vectors shown in figures
\floattable
\begin{deluxetable}{ccccc}
\tabletypesize{\scriptsize}
\colnumbers
\tablecaption{SCUPOL 850~$\mu$m Polarizations\label{tab_SCUPOL}}
\tablewidth{0pt}
\tablehead{
\colhead{Pixel} & \colhead{RA} & \colhead{decl.} & \colhead{$P^\prime$\tablenotemark{a}} & \colhead{$B_{POS}$ PA} \\
&\colhead{[$\degr$]} & \colhead{[$\degr$]} & \colhead{[\%]} &
 \colhead{[$\degr$]} 
}
\startdata
     1 &       76.05390 &  25.18405 & 3.80 (1.21) & 173.1 (9.1) \\
     2 &       76.05390 &  25.19095 & 3.70 (1.20) & 28.8 (9.3) \\
     3 &       76.05390 &  25.19785 & 4.70 (1.92) & 179.7 (11.8) \\
     4 &       76.06152 &  25.17026 & 9.72 (1.56) & 175.2 (4.6) \\
     5 &       76.06152 &  25.17716 & 2.97 (0.74) & 3.0 (7.1) \\
     6 &       76.06152 &  25.18405 & 2.52 (0.61) & 162.6 (7.0) \\
     7 &       76.06152 &  25.19095 & 3.07 (0.75) & 19.2 (7.0) \\
     8 &       76.06152 &  25.19785 & 5.95 (1.38) & 15.6 (6.6) \\
     9 &       76.06914 &  25.17026 & 2.24 (0.75) & 22.0 (9.5) \\
    10 &       76.06914 &  25.17716 & 2.04 (0.45) & 2.24 (6.3) \\
    11 &       76.06914 &  25.18405 & 3.30 (0.46) & 168.3 (4.0) \\
    12 &       76.06914 &  25.19095 & 1.36 (0.70) & 9.8 (14.8) \\
    13 &       76.06914 &  25.19785 & 4.78 (1.53) & 51.8 (9.2) \\
    14 &       76.07676 &  25.16337 & 3.61 (1.15) & 53.8 (9.1) \\
    15 &       76.07676 &  25.17026 & 3.27 (0.73) & 48.0 (6.4) \\
    16 &       76.07676 &  25.17716 & 2.09 (0.54) & 5.4 (7.4) \\
    17 &       76.07676 &  25.18405 & 3.60 (0.66) & 5.0 (5.3) \\
    18 &       76.08438 &  25.17026 & 2.33 (1.16) & 35.2 (14.3) \\
    19 &       76.08438 &  25.17716 & 3.09 (1.18) & 44.5 (11.0) \\
    20 &       76.08438 &  25.18405 & 6.88 (1.43) & 25.2 (6.0) \\
\enddata
\tablenotetext{a}{Pixel values are followed by uncertainties in parentheses.}
\end{deluxetable}

However, examination of Figure~\ref{fig_SCUBA} reveals that the \citet{Matthews09} reprocessing
plus post-processing reveals complexity not seen in the older works.
The Figure shows that the vectors closest to the intensity peak
have a nearly vertical, PA=0$\degr$, orientation while the vectors farther from the
peak have PAs closer to 30-40$\degr$. This distinction can be seen post-facto in
the \citet{WardThompson00} map, though the number of central pixels is only
two or three, and the number of outer pixels is similarly small. The impression
from the reprocessed data 
is that the central $B_{POS}$ orientation is closer to being parallel to the major
axis of the L1544 dense core dust structure, while the outer regions show $B_{POS}$ orientations
more in agreement with the NIR values. Indeed, if the eight new positions located 
closest to the intensity peak are considered as a subset, their mean $B_{POS}$ PA is 
$177.4 \pm 2.2\degr$ and their $\Delta_{PA}$(corrected) is 15.9$\degr$, compared
to PA$ = 22.5 \pm 2.2\degr$ and $\Delta_{PA}$(corrected) of 23.9$\degr$ for
the twelve positions surrounding the core. The mean PAs for the two subsets differ
by 8$\sigma$, indicating significant $B_{POS}$ changes {\it within} the
Arecibo Zeeman beam size. 

% F8
\begin{figure}
\plotone{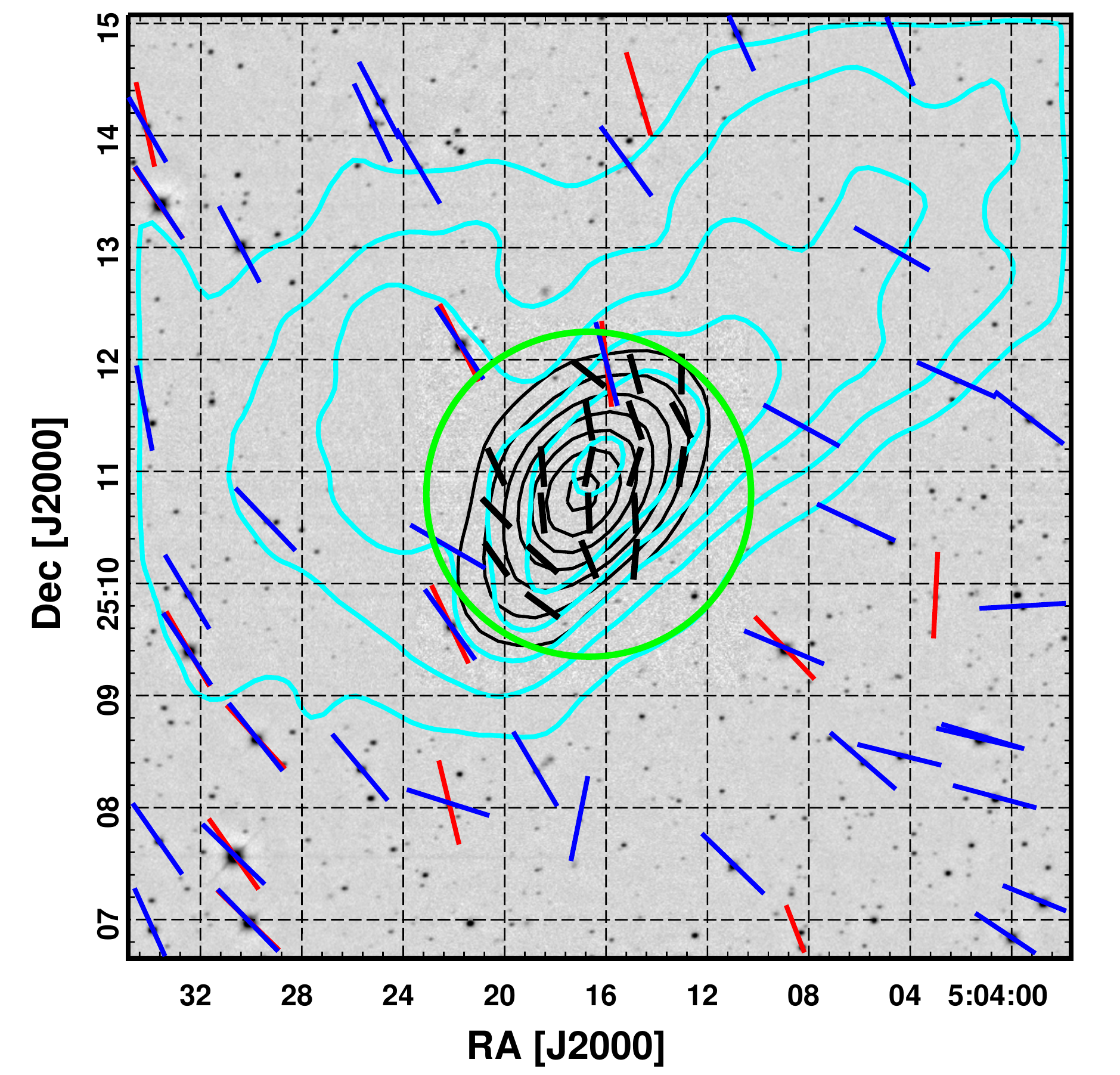}
\caption{Grayscale Mimir deep $H$-band image, zoomed to show the
central $8 \times 8$~arcmin region including the L1544 dense cloud core. Mimir $H$-band polarization
vectors are displayed as blue line segments, with all segment lengths set equal to 
highlight B-field orientations. Mimir $K$-band polarization vectors are similarly shown as red
line segments. Black vectors are the SCUPOL values seen in the previous
Figure. Cyan contours show the {\it Herschel}\, 250~$\mu$m dust
emission. 
Black contours show JCMT/SCUBA Stokes $I$ from the previous Figure. 
The green circle is the Arecibo FWHM beam size for the \citet{Crutcher09} observations.
\label{fig_NIR_SCUBA}}
\end{figure}

A comparison between reprocessed SCUPOL 
and NIR BSP $B_{POS}$ PA values is shown in Figure~\ref{fig_NIR_SCUBA}. This
$8 \times 8$~arcmin Mimir survey region center shows $H$ band
PA vectors in blue, $K$ band PA vectors in red, and (90$\degr$\, rotated)
SCUPOL vectors in black. Together, they reveal a twist in $B_{POS}$ orientations in
going from the dense core out to the larger, lower density region that
starts approximately at the Arecibo beamsize circle, where the SCUPOL
vectors begin to match the NIR vectors. 

\section{Discussion}
\label{sec:discussion}

\subsection{NIR BSP Traces B-fields in L1544}

The evidence presented above shows that the L1544 cloud, as revealed
in the {\it Herschel} 250~$\mu$m thermal dust emission (Figure~\ref{fig_Herschel}), is
well-traced by NIR reddening using the RJCE $E(H-M)$ colors (Figure~\ref{fig_HM_map}).
That same dust is responsible for both BSP in the NIR (Figure~\ref{fig_pol_vecs} and
Figure~\ref{fig_mean_P}) as well as thermal emission polarization in the submm
(Figure~\ref{fig_SCUBA}). 

The {\it changes} in the polarization properties with direction on the sky, especially changes in the 
mean PA orientation (Figure~\ref{fig_mean_PA}) and changes in $\Delta_{PA}$
(Figure~\ref{fig_PA_disp}), correlate strongly with location relative to the `spine' of
the L1544 cloud and its dense core. That the PAs change so dramatically at the
location of the cloud and yet $\Delta_{PA}$ reaches its lowest minimum there
argue for close coupling of the B-field and the gas and dust within L1544.

The strong decrease in the NIR $\Delta_{PA}$ associated with the cloud spine and in the
core, and the overall outer-core agreement of the NIR and SCUBA B-field orientations
(Figure~\ref{fig_NIR_SCUBA}), points to increases in
the B-field strength with gas density in L1544. To quantify this increase, and indeed to
perform a close comparison of $B_{POS}$ to $B_{LOS}$ (radio Zeeman) amplitudes, 
requires establishing the
mean gas density and velocity dispersion across the NIR survey zone and invocation
of the C-F method. These are outside the scope of this current paper, but are the subjects
of later papers in this series.

\subsection{The Non-Uniform B-Field in the L1544 Cloud Envelope}
\label{sec:non_uniform}

The \citet{Crutcher09} analysis of the Zeeman properties of their clouds' envelopes assumed 
that each of the GBT pointings sampled the same, uniform regular B-field and thus measured
a single, representative value of $B_{LOS}$ for each cloud envelope.  
Assessing the validity of this assumption
requires examining whether the GBT beams covered regions of similar or non-similar 
B-fields for each cloud. 
No quantitative measure of B-field uniformity exists to 
easily address when B-fields are uniform enough to return unbiased results when samples
are averaged as per \citet{Crutcher09}. Instead, a statistical
assessment of key polarization properties was performed for L1544  
using the $B_{POS}$ NIR BSP data of Table~\ref{tab_NIR_pols}.

For each of the Arecibo and GBT beams, the $H$ band, $K$ band, and SCUPOL data sets were 
sampled. A suitable gaussian taper was computed for each star or SCUBA position, with respect to 
the center of each Arecibo or GBT beam, with each gaussian taper FWHM width set to the 
corresponding beamsize FWHM width. 
Additional weighting was by the variance of the quantity being `observed' using these synthetic beams. 
The data were selected to be of good quality, by applying $P^\prime$ and PA uncertainty
cuts ($\sigma_P <$ 3\%; $\sigma_{PA} < 45\degr$). Very low uncertainties (associated
with the brightest stars) were trapped 
to be no less than the lowest quartile uncertainties to prevent a few values from dominating the weighted means.

Most of the stellar values used were $H$-band based, which yielded 329 entries after applying the criteria
above. Similarly, $K$ band provided 29 stars and SCUPOL provided all 20 positions. 
The means and uncertainties for polarization PA, $\Delta_{PA}$, and $P^\prime$
were computed for the different combination of $H$ band and $K$ band samples, as well as
$H$+$K$ bands, SCUPOL, and SCUPOL+$H$+$K$ 
sample combinations. Note that SCUPOL points 
cover only the central, Arecibo beam.

\begin{figure}
\plotone{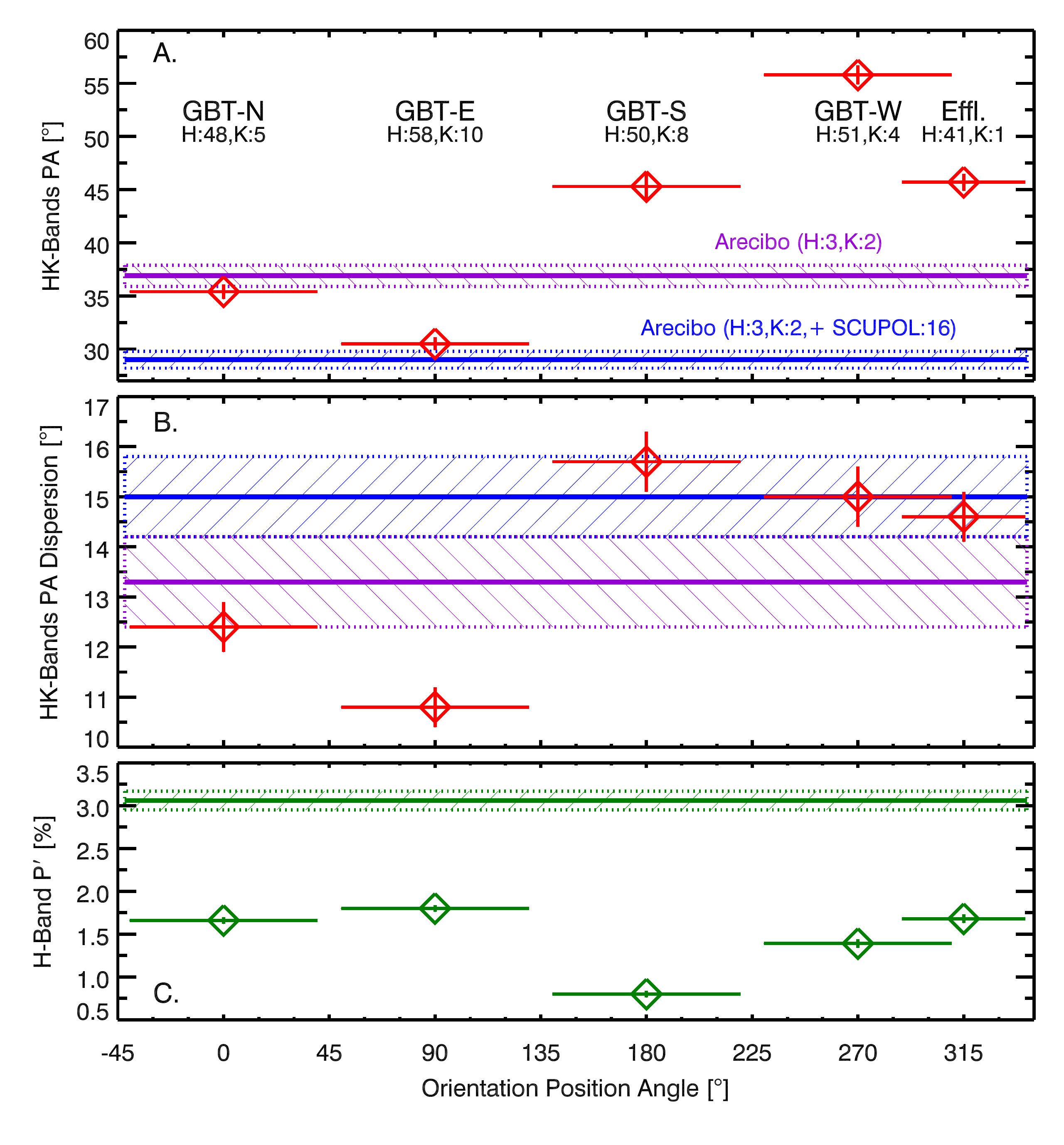}
\caption{Comparison of $H$, $K$, and SCUPOL 
polarizations in the \citet{Crutcher09} Arecibo and GBT beams (as well as in the 
Effelsberg beam). Horizontal
axis is Orientation PA of the GBT beams, seen from 
Arecibo beam center. Red and green diamonds with error bars are  
beam-averaged values, uncertainties, and Orientation PAs. Top, A, 
panel shows polarization PA, from
$H$ and $K$. Beam designations and effective stellar numbers by 
band (see text) run along panel top. PAs for Arecibo are violet horizontal
line and hatching for $H$ and $K$, and blue line
and hatching when also including SCUPOL data. Middle, B, panel  
compares $\Delta_{PA}$. Bottom, C, panel
compares $H$-band $P^\prime$, with
Arecibo value and uncertainty as the green line and hatching. Strong
beam-to-beam variations in PA and $P^\prime$ and weaker variations in
$\Delta_{PA}$ indicate $B_{POS}$ is 
unlikely to be
highly uniform in the GBT-sampled 
L1544 envelope.
\label{fig_PA_vs_PA}}
\end{figure}

Figure~\ref{fig_PA_vs_PA} shows, and Table~\ref{tbl_beams} lists, the beam-based 
comparisons of plane-of-sky means of PA, $\Delta_{PA}$, and $P^\prime$ values. 
In the Figure, the $x$-axis displays `Orientation PA,' 
which was defined as the projected 
PA from the Arecibo beam center to each of the regions covered by a GBT beam 
(with the usual East-from-North angle increment). 
Thus, the `GBT-N' (North) beam is centered at Orientation PA=0$\degr$, but spans about 
$\pm 40\degr$\, of Orientation PA. 
In this plot, a uniform B-field in the envelope of L1544 would show the same polarization PA 
(or other property) for all Orientation PAs. 
The Table presents the NIR weighted means and uncertainties, using $H$ and $K$ band stellar
data for PA and $\Delta_{PA}$ and $H$ band data alone for $P^\prime$. For the Arecibo beam
average, the first line in the Table presents the $H$ and $K$ values, while the second line
includes the effects of the (weighted) SCUPOL points. 

In the top panel of Figure~\ref{fig_PA_vs_PA}, the red points show the polarization PA 
averages and uncertainties as well as the effective widths of the GBT beams in Orientation PA.
The numbers below each GBT beam identifier list the effective number of stars used in the 
H and K band averages. These are effective numbers because of the gaussian tapers - all of the 
individual stars contribute, but only in summed gaussian weights equivalent to the listed numbers. 
The violet line and 
hatching show the same polarization PA and uncertainty for the H and K stars in the Arecibo 
beam (much smaller than the GBT beams and centered on the opaque cloud core). The 
blue line and hatching add the SCUPOL values to the H and K ones. The Orientation PA of the 
Arecibo beam spans 0-360$\degr$, hence the use of the hatched regions to render its values.

This panel displays just how different the 
polarization properties are in the different GBT and Effelsberg beams. While the 
polarization PAs of two of the beams (GTB-N and GBT-E) are partially consistent with each other (differing by 5$\sigma$) 
and also with one, or the other, Arecibo beam estimate, the other two beams' polarization 
PAs are quite different (6.5 sigma for GBT-S and 23 sigma for GBT-W, both compared to 
GBT-N). These position-to-position differences indicate that the B-field is unlikely to be uniform in the envelope surrounding the L1544 core.

The middle panel is the same type of comparison for $\Delta_{PA}$ values.
The differences, compared to the values for the central Arecibo beam, are less significant 
here, though large differences between the values in the GBT beams remain. This
is especially true when comparing the GBT-N and -E beam values to the GBT-S and -W beam
values.

% Fig 9

% NIR BSP in the Zeeman Beams
\floattable
\begin{deluxetable}{ccccc}
\tabletypesize{\scriptsize}
\tablewidth{0cm}
\tablecaption{NIR BSP Properties in Arecibo, GBT, and Effelsberg Beams\label{tbl_beams}}
\tablehead{
\colhead{Region} & \colhead{Orientation PA} & \colhead{$<P^\prime_H>$\tablenotemark{a}} & \colhead{$<PA>$} & \colhead{$\sigma_{PA}$} \\
& \colhead{[$\degr$]}  & \colhead{[\%]} & \colhead{[$\degr$]} & \colhead{[$\degr$]}  \\
\colhead{(1)} & \colhead{(2)} & \colhead{(3)} & \colhead{(4)} & \colhead{(5)}
}
\startdata
Arecibo - Dense Core - HK only	& ... & 3.06 (0.11) & 36.9 (1.0) & 13.3 (0.9)  \\
Arecibo - HK+SCUPOL & ... & ... & 29.0 (0.8) & 15.0 (0.8) \\ [6pt]
GBT - North &  $-$40 - $+$40 & 1.66 (0.04) & 35.4 (0.7) & 12.4 (0.5) \\
GBT - East  & 50 - 130 & 1.80 (0.04) & 30.5 (0.6) & 10.8 (0.4)  \\
GBT - South & 140 - 220 & 0.80 (0.04) & 45.3 (1.3) & 15.7 (0.6)  \\
GBT - West & 230 - 310 & 1.39 (0.05) & 55.8 (0.9) & 15.0 (0.6)  \\
Effelsberg - NW & 289 - 341 & 1.68 (0.05) & 45.7 (0.8) & 14.6 (0.5)  \\
\enddata
\tablenotetext{a}{Average values are followed by uncertainties in parentheses.}
\end{deluxetable}

The bottom panel compares $P^\prime$ values for $H$ band, only,
as no other similar comparison covers all of the beams.
The GBT beams would not be expected to contain background stars exhibiting polarizations 
as high as those seen in the Arecibo beam, but a uniform B-field in the L1544 envelope might be 
expected to yield better uniformity of $P^\prime$ across the GBT beams. This is not what is seen here:
the GBT-N and GBT-S beams show $P^\prime$ values that differ by 15$\sigma$. 

The statistically different PA orientations and dispersions in the L1544 envelope-tracing
GBT beams, and indeed the PA twist newly uncovered in the SCUPOL data within the
Arecibo beam, 
suggest that simple averaging of Zeeman detections (especially with non-detections)
for the purpose of estimating relative core-envelope $M/\Phi$ values by \citet{Crutcher09} 
is likely to be biased. 
Detailed conclusions regarding the applicability of 
AD models to dense core formation and evolution, which rest on 
relative core/envelope $M/\Phi$ estimates,
must be revisited using more robust observational approaches, including deeper analyses
of the current NIR BSP data. 

\section{Summary}

Accurately characterizing B-fields is challenging, but is vital to understanding how molecular
clouds form, evolve, and produce new stars. Testing leading B-field models that address these
phases is equally important, and must be performed using a variety of techniques and tools.
Here, new near-infrared imaging stellar background polarimetry, and post-processing of the
archived re-reduced SCUPOL data, were used to survey the full extent of the
L1544 dark cloud, which has the best radio OH Zeeman effect detections (of its core and
one envelope position) of any dark cloud.

The first goal of this study was to show that near-infrared starlight polarimetry is able
to reveal B-fields across the L1544 cloud at high angular sampling and precision. This goal 
was met by
revealing that the positional changes in plane-of-sky B-field PA orientations and dispersions 
of those orientations correlate strongly with the location and structure of the L1544 dust
thermal emission.

The second goal was to test whether the plane-of-sky polarization properties were 
uniform throughout the envelope of L1544 or whether these properties vary significantly.
A key assumption of the analysis of OH Zeeman observations, performed by 
\citet{Crutcher09} and leading to relative mass-to-flux ratios of the L1544 core and envelope,
was that the B-field was uniform in the envelope, permitting averaging of the Zeeman
observations across the four GBT beam pointings without accounting for possible beam-to-beam intrinsic variations.

The near-infrared background starlight polarimetry, averaged over each of the different
GBT Zeeman beam sizes and positions observed, instead showed strong beam-to-beam variations 
in the plane-of-sky polarization properties. The reprocessed SCUPOL
data showed a similar strong change in PA directions within the much smaller Arecibo
beamsize used for the initial Zeeman detection of the core B-field.

Averaging low-signal Zeeman observations from different pointings 
without treating intrinsic variations would be effective if the B-field were uniform across the pointings.
For L1544, the near-infrared polarimetry results are at odds with this
uniformity assumption and thereby the conclusions that rest upon it. 

\acknowledgments

The authors thank Dick Crutcher and Tom Troland for answering questions concerning
their Zeeman observations of L1544 and thank the two reviewers for their sincere efforts to 
improve the paper. Brian Taylor rebuilt a key Mimir computer just in
time to enable the 2016 January observations on the last clear night of the run.
This publication makes use of data products from the Two Micron All Sky Survey, 
which is a joint project of the University of Massachusetts and the Infrared 
Processing and Analysis Center/California Institute of Technology (CalTech), funded by 
NASA and NSF. 
AllWISE makes use of data from {\it WISE}, which is a joint project of the University 
of California, Los Angeles, and the Jet Propulsion Laboratory (JPL)/CalTech, and 
NEOWISE, which is a project of JPL/CalTech. WISE and NEOWISE are funded 
by NASA.
This research has made use of the VizieR catalogue access tool, CDS,
Strasbourg, France. VizieR is a joint effort of CDS (Centre de Données
astronomiques de Strasbourg) and ESA-ESRIN (Information Systems
Division).

This research was conducted in part at JPL, which is
operated for NASA by CalTech.
K.T. acknowledges support by FP7 through Marie Curie Career Integration Grant 
PCIG-GA-2011-293531 ``SFOnset," and partial support from the EU FP7 Grant 
PIRSES-GA-2012-31578 ``EuroCal."
This research was conducted in part 
using the Mimir instrument, jointly developed at Boston University (BU) and Lowell 
Observatory (LO) and supported by NASA, NSF, and the W.M. Keck Foundation.
Mimir observations and analyses have been made possible by grants 
AST 06-07500, 09-07790, and 14-12269 from NSF/MPS, by 
NASA ADAP grant NNX15AE51G, 
and by grants of significant observing time from the BU - LO 
partnership. 

\facilities{Perkins, {\it Herschel}, {\it WISE}, JCMT, GBT, Arecibo}
\software{IDL, DS9}


\begin{thebibliography}{}

\bibitem[Arce et al.(1998)]{Arce98} Arce, H.~G., Goodman, 
A.~A., Bastien, P., Manset, N., \& Sumner, M.\ 1998, \apjl, 499, L93 

\bibitem[Bacmann et al.(2000)]{Bacmann00} Bacmann, A., Andr{\'e}, P., Puget, J.-L., et al.\ 2000, \aap, 361, 555 

\bibitem[Basu \& Mouschovias(1994)]{Basu94} Basu, S., \& Mouschovias, T.~Ch.\ 1994, \apj, 432, 720

\bibitem[Benson et al.(1998)]{Benson98} Benson, P. J., Caselli, P., \& Myers, P. C.\ 1998, \apj, 506, 743

\bibitem[Bizzocchi et al.(2014)]{Bizzocchi14} Bizzocchi, L., Caselli, P., Spezzano, S., \& Leonardo, E.\ 2014, \aap, 569, A27 

\bibitem[Caselli et al.(2002a)]{Caselli02a} Caselli, P., Walmsley, C. M., Zucconi, A., et al.\ 2002a, \apj, 565, 331

\bibitem[Caselli et al.(2002b)]{Caselli02b} Caselli, P., Walmsley, 
C.~M., Zucconi, A., et al.\ 2002b, \apj, 565, 344 

\bibitem[Caselli et al.(2002c)]{Caselli02c} Caselli, P., Benson, 
P.~J., Myers, P.~C., \& Tafalla, M.\ 2002c, \apj, 572, 238 

\bibitem[Caselli et al.(2010)]{Caselli10} Caselli, P., Keto, E., Pagani, L., et al.\ 2010, \aap, 521, L29 

\bibitem[Chandrasekhar \& Fermi(1953)]{CF53} Chandrasekhar, S., \& Fermi, E.\ 1953, \apj, 118, 113 

\bibitem[Ciolek \& Mouschovias(1994)]{Ciolek94} Ciolek, G.~E., \& Mouschovias, T.~Ch.\ 1994, \apj, 425, 142 

\bibitem[Ciolek \& Basu(2000)]{Ciolek00} Ciolek, G.~E., \& Basu, S.\ 2000, \apj, 529, 925 

\bibitem[Clemens et al.(2007)]{Clemens07} Clemens, D.~P., Sarcia, 
D., Grabau, A., et al.\ 2007, \pasp, 119, 1385 

\bibitem[Clemens et al.(2012c)]{Clemens12c} Clemens, D.~P., Pavel, 
M.~D., \& Cashman, L.~R.\ 2012c, \apjs, 200, 21 

\bibitem[Clemens et al.(2012b)]{Clemens12b} Clemens, D.~P., 
Pinnick, A.~F., \& Pavel, M.~D.\ 2012b, \apjs, 200, 20 

\bibitem[Clemens et al.(2012a)]{Clemens12a} Clemens, D.~P., 
Pinnick, A.~F., Pavel, M.~D., \& Taylor, B.~W.\ 2012a, \apjs, 200, 19 

\bibitem[Crapsi et al.(2005)]{Crapsi05} Crapsi, A., Caselli, P., 
Walmsley, C.~M., et al.\ 2005, \apj, 619, 379 

\bibitem[Crutcher(1999)]{Crutcher99} Crutcher, R.~M.\ 1999, \apj, 520, 706

\bibitem[Crutcher(2012)]{Crutcher12} Crutcher, R.~M.\ 2012, \araa, 50, 29 

\bibitem[Crutcher et al.(1996)]{Crutcher96} Crutcher, R.~M., 
Troland, T.~H., Lazareff, B., \& Kazes, I.\ 1996, \apj, 456, 217 

\bibitem[Crutcher et al.(2009)]{Crutcher09} Crutcher, R. M., Hakobkian, N., \& Troland, T. H.\ 2009, \apj, 692, 844

\bibitem[Crutcher \& Troland(2000)]{Crutcher00} Crutcher, R. M., \& Troland, T. H.\ 2000, \apj, 537, L139

\bibitem[Crutcher et al.(2004)]{Crutcher04} Crutcher, R.~M., 
Nutter, D.~J., Ward-Thompson, D., \& Kirk, J.~M.\ 2004, \apj, 600, 279 

\bibitem[Cutri et al.(2003)]{Cutri03} Cutri, R.~M., Skrutskie, M.~F., van Dyk, S., et al.\ 2003, VizieR Online Data Catalog, 2246, 0

\bibitem[Cutri et al.(2013)]{Cutri13} Cutri, R.~M., \& et al.\ 2013, VizieR Online Data Catalog, 2328, 0 

\bibitem[Doty et al.(2005)]{Doty05} Doty, S.~D., Everett, 
S.~E., Shirley, Y.~L., Evans, N.~J., 
\& Palotti, M.~L.\ 2005, \mnras, 359, 228 

\bibitem[Elias(1978)]{Elias78} Elias, J.~H.\ 1978, \apj, 224, 
857 

\bibitem[Fiedler \& Mouschovias(1992)]{Fiedler92} Fiedler, R.~A., \& Mouschovias, T.~Ch.\ 1992, \apj, 391, 199

\bibitem[Fiege \& Pudritz(2000)]{Fiege00} Fiege, J.~D., \& Pudritz, R.~E.\ 2000, \mnras, 311, 85 

\bibitem[Galli \& Shu(1993)]{Galli93} Galli, D., \& Shu, F.~H.\ 1993, \apj, 417, 220 

\bibitem[Girart et al.(1999)]{Girart99} Girart, J.~M., Crutcher, 
R.~M., \& Rao, R.\ 1999, \apjl, 525, L109 

\bibitem[Goldreich \& Kylafis(1981)]{Goldreich81} Goldreich, P., \& Kylafis, N.~D.\ 1981, \apjl, 243, L75 

\bibitem[Goodman et al.(1989)]{Goodman89} Goodman, A.~A., Crutcher, R.~M., Heiles, C., Myers, P.~C., \& Troland, T.~H.\ 1989, \apjl, 338, L61

\bibitem[Goodman et al.(1995)]{Goodman95} Goodman, A.~A., Jones, T.~J., Lada, E.~A., \& Myers, P.~C.\ 1995, \apj, 448, 748 

\bibitem[Greaves et al.(2003)]{Greaves03} Greaves, J.~S., Holland, W.~S., Jenness, T., et al.\ 2003, \mnras, 340, 353 

\bibitem[Hennebelle \& Fromang(2008)]{Hennebelle08} Hennebelle, P., \& Fromang, S.\ 2008, \aap, 477, 9 

\bibitem[Heyer et al.(1987)]{Heyer87} Heyer, M.~H., Vrba, 
F.~J., Snell, R.~L., et al.\ 1987, \apj, 321, 855 

\bibitem[Holland et al.(1999)]{Holland99} Holland, W.~S., Robson, E.~I., Gear, W.~K., et al.\ 1999, \mnras, 303, 659 

\bibitem[Jones et al.(2015)]{Jones15} Jones, T.~J., Bagley, M., 
Krejny, M., Andersson, B.-G., \& Bastien, P.\ 2015, \aj, 149, 31 

\bibitem[Joye \& Mandel(2003)]{Joye03}Joye, W. A., \& Mandel, E. 2003, 
in Astronomical Data Analysis Software and Systems XII, 
H. E. Payne, R. I. Jedrzejewski, \& R. N. Hook, eds.,
ASP Conf. Ser., 295, 489

\bibitem[Kenyon et al.(1994)]{Kenyon94} Kenyon, S.~J., 
Dobrzycka, D., \& Hartmann, L.\ 1994, \aj, 108, 1872 

\bibitem[Keto \& Caselli(2008)]{Keto08} Keto, E., \& Caselli, P.\ 2008, \apj, 683, 238 

\bibitem[Keto \& Caselli(2010)]{Keto10} Keto, E., \& Caselli, P.\ 2010, \mnras, 402, 1625 

\bibitem[Keto et al.(2014)]{Keto14} Keto, E., Rawlings, J., 
\& Caselli, P.\ 2014, \mnras, 440, 2616 

\bibitem[Keto et al.(2015)]{Keto15} Keto, E., Caselli, P., 
\& Rawlings, J.\ 2015, \mnras, 446, 3731 

\bibitem[Lee et al.(2003)]{Lee03} Lee, J.-E., Evans, N.~J., 
II, Shirley, Y.~L., \& Tatematsu, K.\ 2003, \apj, 583, 789 

\bibitem[Li et al.(2014)]{Li14} Li, H.-B., Goodman, A., Sridharan, T.~K., et al.\ 2014, Protostars and Planets VI, 101 

\bibitem[Li et al.(2015)]{Li15} Li, P.~S., McKee, C.~F., \& Klein, R.~I.\ 2015, \mnras, 452, 2500 

\bibitem[Li \& Nakamura(2002)]{Li02a} Li, Z.-Y., \& Nakamura, F.\ 2002, \apj, 578, 256 

\bibitem[Li et al.(2002)]{Li02b} Li, Z.-Y., Shematovich, 
V.~I., Wiebe, D.~S., \& Shustov, B.~M.\ 2002, \apj, 569, 792 

\bibitem[Majewski et al.(2011)]{Majewski11} Majewski, S.~R., 
Zasowski, G., \& Nidever, D.~L.\ 2011, \apj, 739, 25 

\bibitem[Matthews et al.(2009)]{Matthews09} Matthews, B.~C., 
McPhee, C.~A., Fissel, L.~M., \& Curran, R.~L.\ 2009, \apjs, 182, 143 

\bibitem[Messinger et al.(1997)]{Messinger97} Messinger, D.~W., Whittet, D.~C.~B.,
\& Roberge, W.~G.\ 1997, \apj, 487, 314

\bibitem[Mestel \& Spitzer(1956)]{Mestel56} Mestel, L., \& Spitzer, L., Jr.\ 1956, \mnras, 116, 503 

\bibitem[Mouschovias(1976a)]{Mouschovias76a} Mouschovias, T.~Ch.\ 1976a, \apj, 206, 753 

\bibitem[Mouschovias(1976b)]{Mouschovias76b} Mouschovias, T.~Ch.\ 1976b, \apj, 207, 141 

\bibitem[Mouschovias(1996a)]{Mouschovias96a} Mouschovias, T.~Ch.\ 1996a, NATO Advanced Science Institutes (ASI) Series C, 481, 475

\bibitem[Mouschovias(1996b)]{Mouschovias96b} Mouschovias, T.~Ch.\ 1996b, NATO Advanced Science Institutes (ASI) Series C, 481, 505

\bibitem[Mouschovias(1991)]{Mouschovias91} Mouschovias, T.~Ch.\ 1991, \apj, 373, 169 

\bibitem[Mouschovias et al.(2006)]{Mouschovias06} Mouschovias, T.~Ch., Tassis, K., \& Kunz, M.~W.\ 2006, \apj, 646, 1043 

\bibitem[Mouschovias \& Tassis(2010)]{Mouschovias10} Mouschovias, T.~Ch., \&
Tassis, K.\ 2010, \mnras, 409, 801

\bibitem[Myers et al.(1983)]{Myers83a} Myers, P.~C., Linke, 
R.~A., \& Benson, P.~J.\ 1983, \apj, 264, 517 

\bibitem[Myers \& Benson(1983)]{Myers83b} Myers, P.~C., \& Benson, P.~J.\ 1983, \apj, 266, 309 

\bibitem[Padoan \& Nordlund(1999)]{Padoan99} Padoan, P., \& Nordlund, {\AA}.\ 1999, \apj, 526, 279 

\bibitem[Pillai et al.(2015)]{Pillai15} Pillai, T., Kauffmann, J., Tan, J.~C., et al.\ 2015, \apj, 799, 74 

\bibitem[Serkowski et al.(1975)]{Serkowski75} Serkowski, K., Mathewson, D.S., \& Ford, V.L. 1975,
ApJ, 196, 261

\bibitem[Shirley et al.(2000)]{Shirley00} Shirley, Y.~L., Evans, 
N.~J., II, Rawlings, J.~M.~C., \& Gregersen, E.~M.\ 2000, \apjs, 131, 249 

\bibitem[Skrutskie et al.(2006)]{Skrutskie06} Skrutskie, M.F., Cutri, R. M., Stiening, R.,
et al. 2006, AJ, 131, 1163.

\bibitem[Snell(1981)]{Snell81} Snell, R.~L.\ 1981, \apjs, 45, 121 

\bibitem[Tafalla et al.(1998)]{Tafalla98} Tafalla, M., Mardones, 
D., Myers, P.~C., et al.\ 1998, \apj, 504, 900 

\bibitem[Tafalla et al.(2002)]{Tafalla02} Tafalla, M., Myers, 
P.~C., Caselli, P., Walmsley, C.~M., \& Comito, C.\ 2002, \apj, 569, 815 

\bibitem[Torres et al.(2012)]{Torres12} Torres, R.~M., Loinard, 
L., Mioduszewski, A.~J., et al.\ 2012, \apj, 747, 18 

\bibitem[Troland \& Crutcher(2008)]{Troland08} Troland, T. H., \& Crutcher, R. M.\ 2008, \apj, 680, 457

\bibitem[van der Tak et al.(2005)]{vanderTak05} van der Tak, F.~F.~S., Caselli, P., \& Ceccarelli, C.\ 2005, \aap, 439, 195 

\bibitem[Vastel et al.(2006)]{Vastel06} Vastel, C., Caselli, P., 
Ceccarelli, C., et al.\ 2006, \apj, 645, 1198 

\bibitem[Ward-Thompson et al.(1999)]{WardThompson99} Ward-Thompson, 
D., Motte, F., \& Andre, P.\ 1999, \mnras, 305, 143 

\bibitem[Ward-Thompson et al.(2000)]{WardThompson00} Ward-Thompson, 
D., Kirk, J.~M., Crutcher, R.~M., et al.\ 2000, \apjl, 537, L135 

\bibitem[Wardle \& Kronberg(1974)]{Wardle1974} Wardle, J.~F.~C., \& Kronberg, P.~P.\ 1974, \apj, 194, 249 

\bibitem[Whittet et al.(2008)]{Whittet08} Whittet, D.~C.~B., 
Hough, J.~H., Lazarian, A., \& Hoang, T.\ 2008, \apj, 674, 304 

\bibitem[Wilking et al.(1980)]{Wilking80}Wilking, B. A., Lebofsky, M. J., Kemp, J. C., et al. 1980, ApJ, 235, 905

\bibitem[Williams et al.(1999)]{Williams99} Williams, J.~P., 
Myers, P.~C., Wilner, D.~J., \& Di Francesco, J.\ 1999, \apjl, 513, L61 

\bibitem[Williams et al.(2006)]{Williams06} Williams, J.~P., Lee, 
C.~W., \& Myers, P.~C.\ 2006, \apj, 636, 952 

\bibitem[Wright et al.(2010)]{Wright10} Wright, E.~L., 
Eisenhardt, P.~R.~M., Mainzer, A.~K., et al.\ 2010, \aj, 140, 1868 

\bibitem[Zacharias et al.(2010)]{Zacharias10} Zacharias, N., Finch, C., Girard, T., et al.\ 2010, \aj, 139, 2184 

\bibitem[Zhang et al.(2014)]{Zhang14} Zhang, Q., Qiu, K., Girart, J.~M., et al.\ 2014, \apj, 792, 116 

\end{thebibliography}
\end{document}